\documentclass[nofootinbib,a4paper,12pt]{article}
\usepackage{jheppub}
\usepackage[T1]{fontenc}
\usepackage[utf8]{inputenc}
\usepackage{float}
\usepackage{slashed}
\usepackage{multicol,multirow}
\usepackage{amssymb}
\usepackage{amsmath}
\usepackage{subcaption}
\usepackage{array}
\usepackage{xcolor}
\usepackage{hyperref}
\hypersetup{colorlinks=true}
\usepackage{longtable}
\usepackage{graphics,graphicx,epsfig,ulem,booktabs}
\numberwithin{equation}{section}
\usepackage{url}

\definecolor{green}{rgb}{0.1,0.8,0.2}
\definecolor{orange}{rgb}{1.0,0.5,0.0}
\definecolor{cyan}{rgb}{0.0,0.75,0.8}
\definecolor{brown}{rgb}{0.7,0.35,0.05}

\newcolumntype{C}[1]{>{\centering\let\newline\\\arraybackslash\hspace{0pt}}m{#1}}

\newcommand{\Gu}{\Gamma (B^+)}
\newcommand{\Gd}{\Gamma (B_d)}
\newcommand{\Gs}{\Gamma (B_s)}
\newcommand{\tud}{\tau (B^+)/\tau (B_d)}
\newcommand{\tsd}{\tau (B_s)/\tau (B_d)}
\newcommand{\GeV}{\, {\rm GeV}}
\newcommand{\psinv}{\, {\rm ps}^{-1}}
\newcommand{\mupi}{\mu_\pi^2 (B_q)}
\newcommand{\muG}{\mu_G^2 (B_q)}
\newcommand{\rhoD}{\rho_D^3 (B_q)}
\newcommand{\mupis}{\mu_\pi^2 (B_s)}
\newcommand{\muGs}{\mu_G^2 (B_s)}
\newcommand{\rhoDs}{\rho_D^3 (B_s)}

\title{\boldmath Total decay rates of $B$ mesons at NNLO-QCD}

\preprint{\small TUM-HEP-1545/24, P3H-24-101, SI-HEP-2024-31, TTP24-046, Nikhef~2024-019, ZU-TH~67/24}

\author[a]{Manuel Egner,}
\author[b,c]{Matteo Fael,}
\author[d]{Alexander Lenz,}
\author[d,e,f]{Maria Laura Piscopo,}
\author[d,g]{Aleksey~V.~Rusov,}
\author[h]{Kay Schönwald,}
\author[a]{Matthias Steinhauser}

\affiliation[a]{Institut für Theoretische Teilchenphysik, Karlsruhe Institute of Technology (KIT), Wolfgang-Gaede Straße 1, 76131 Karlsruhe, Germany}
\affiliation[b]{Dipartimento di Fisica e Astronomia ``G. Galilei'', 
Università di Padova, Via F.\ Marzolo 8, 35131 Padova, Italy}
\affiliation[c]{Istituto Nazionale di Fisica Nucleare, Sezione di Padova, Via F.\ Marzolo 8, 35131 Padova, Italy}
\affiliation[d]{Physik Department, Universit\"{a}t Siegen, Walter-Flex-Str. 3, 57068 Siegen, Germany}
\affiliation[e]{Nikhef, Science Park 105, NL-1098 XG Amsterdam, Netherlands}
\affiliation[f]{Department of Physics and Astronomy, Vrije Universiteit Amsterdam, NL-1081 HV Amsterdam, Netherlands}
\affiliation[g]{Physik Department T31, 
Technische Universit\"at M\"unchen, James-Franck-Straße 1, D–85748 Garching, Germany}
\affiliation[h]{Physik-Institut, Universität Zürich, Winterthurerstrasse 190, 8057 Zürich, Switzerland}

\emailAdd{manuel.egner@kit.edu}
\emailAdd{matteo.fael@pd.infn.it}
\emailAdd{alexander.lenz@uni-siegen.de}
\emailAdd{mpiscopo@nikhef.nl}
\emailAdd{aleksey.rusov@tum.de}
\emailAdd{kay.schoenwald@physik.uzh.ch}
\emailAdd{matthias.steinhauser@kit.edu}

\abstract{We update the Standard Model (SM) predictions for the lifetimes of the $B^+$, $B_d$ and $B_s$ mesons 
within the heavy quark expansion (HQE),
including the recently determined NNLO-QCD corrections to non-leptonic decays of the free $b$-quark. In addition, we update the HQE predictions for the lifetime ratios $\tau (B^+)/\tau (B_d)$ and $\tau (B_s)/\tau (B_d)$, and provide new results for the semileptonic branching fractions of the three mesons entirely within the HQE. We obtain a considerable improvement of the theoretical uncertainties, mostly due to the reduction of the renormalisation scale dependence when going from LO to NNLO, and for all the observables considered, we find good agreement, within uncertainties, between the HQE predictions and the corresponding experimental data.
Our results read, respectively,
$
\Gamma (B^+) = 0.587^{+0.025}_{-0.035}~{\rm ps}^{-1},\,
\Gamma (B_d) = 0.636^{+0.028}_{-0.037}~{\rm ps}^{-1},\,
\Gamma (B_s) = 0.628^{+0.027}_{-0.035}~{\rm ps}^{-1},\,
$
for the total decay widths,
$
\tau (B^+)/\tau (B_d) = 1.081^{+0.014}_{-0.016}, \,
\tau (B_s)/\tau (B_d) = 1.013^{+0.007}_{-0.007}, \,
$
for the lifetime ratios,
and 
$
{\cal B}_{\rm sl} (B^+) = (11.46^{+0.47}_{-0.32}) \%, \,
{\cal B}_{\rm sl} (B_d) =  (10.57^{+0.47}_{-0.27}) \%, \,
{\cal B}_{\rm sl} (B_s) =  (10.52^{+0.50}_{-0.29}) \%,
$ 
for the semileptonic branching ratios.
Finally, we also provide an outlook for further improvements of the HQE determinations of the $B$-meson decay widths and of their ratios.
}

\begin{document}

\maketitle
\flushbottom

\section{Introduction}
Lifetimes of weakly decaying $B_q$ mesons with a light spectator quark $q = u,d,s$, are by now measured with a precision of per-mille, see the web-update of Ref.~\cite{PDG:2024}:\footnote{Based on the measurements in Refs.~\cite{DELPHI:2003hqy,ALEPH:2000kte,ALEPH:1996geg,DELPHI:1995hxy,DELPHI:1995pkz,DELPHI:1996dkh,L3:1998pnf,OPAL:1995bfe,OPAL:1998msi,OPAL:2000qeg,SLD:1997wak,CDF:1998pvs,CDF:2002ixx,CDF:2010ibe,D0:2008nly,D0:2014ycx,BaBar:2001mmd,BaBar:2002nat,BaBar:2002jxa,BaBar:2002war,BaBar:2005laz,Belle:2004hwe,
CMS:2017ygm,
LHCb:2014qsd,
LHCb:2014bqh,
CDF:2010gif,
D0:2004ije,
ALEPH:1997rqk,
DELPHI:2000aij,
OPAL:1997zgk,CDF:1998htf,DELPHI:2000gjz,OPAL:1997ufs,CDF:2011utg,LHCb:2013cca,LHCb:2014wet,LHCb:2017knt,CDF:1997axv,D0:2004jzq,LHCb:2021awg,CMS:2019bbr,ALEPH:2000cjd,LHCb:2012zwr,LHCb:2013dzm,CDF:2011kjt,D0:2016nbv,LHCb:2016crj,LHCb:2013odx,CDF:2012nqr,D0:2011ymu,ATLAS:2014nmm,ATLAS:2016pno,ATLAS:2020lbz,CMS:2015asi,CMS:2020efq,LHCb:2014iah,LHCb:2017hbp,LHCb:2016tuh,
 LHCb:2019sgv,LHCb:2019nin,LHCb:2021wte}.
Note that very recently the ATLAS collaboration presented a new very precise measurement of the $B_d$-meson lifetime, namely $\tau (B_d) = (1.5053 \pm 0.0012 \pm 0.0035) \, {\rm ps}$ \cite{ATLAS:2024oki}, which is not yet included in the HFLAV average.}
\begin{equation}
\tau (B_d)|_{\rm exp.} = 1.517(4) \,  {\rm ps} \, , 
\quad
\tau (B^+)|_{\rm exp.} = 1.638(4) \,  {\rm ps} \, ,
\quad
\tau (B_s)|_{\rm exp.} = 1.520(5) \,  {\rm ps} \, .
\label{eq:exp-lifetimes}
\end{equation}
Theoretically, the $B_q$ total decay rate, $\Gamma_{B_q} = 1/ \tau (B_q)$, can be systematically computed in the framework of the heavy quark expansion (HQE) as an expansion in inverse powers of the heavy $b$-quark mass $m_b$~\cite{Shifman:1986mx} -- see also the review \cite{Lenz:2014jha} for details on the historical development of the HQE. The leading contribution to the HQE is given by the decay rate of a free heavy $b$ quark $\Gamma_b$, which is universal for each decaying $B_q$ meson. The subleading contributions $\delta \Gamma_{B_q}$, on the other hand, are specific to the $B_q$ meson considered and suppressed by at least two powers of $m_b$. This reads, schematically
\begin{equation}
  \Gamma (B_q) = \Gamma_b + \delta \Gamma_{B_q}\,,
  \qquad
  \delta \Gamma_{B_q} = {\cal O} 
  \left(\frac{1}{m_b^2}\right) \,.
\end{equation}
As the free $b$-quark decay is proportional to the factor 
\begin{equation}
\Gamma_0
= \frac{G_F^2 m_b^5 |V_{cb}|^2}{192 \pi^3}
\, ,
\label{eq:Gamma_0}
\end{equation}
$\Gamma_b$ shows a strong dependence on
the value of the mass of the $b$ quark, leading to large uncertainties, particularly at LO-QCD, where the definition of the quark mass is not fixed.
In ratios of lifetimes of $B_q$ mesons, however, the dependence on the free $b$-quark decay contribution can be removed. In fact, starting from
\begin{eqnarray}
  \frac{\tau(B_q)}{\tau(B_{q^\prime})} & = & \frac{\Gamma_b + \delta \Gamma_{B_{q^\prime}}}{\Gamma_b + \delta \Gamma_{B_{q}}} =
  1 + \left( \delta \Gamma_{B_{q^\prime}} - \delta \Gamma_{B_q}\right)  \tau(B_q) \, ,
  \label{eq:tau_ratio}
\end{eqnarray}
and combining the HQE result for
$(\delta \Gamma_{B_{q^\prime}} - \delta \Gamma_{B_q} )$ with the experimental value of $\tau(B_q)$, it is possible to obtain a prediction for the lifetime ratio which is independent of
$\Gamma_b$ and therefore only sensitive to subleading HQE corrections.
Experimentally these ratios have also been determined with a precision of per-mille~\cite{PDG:2024}: 
\begin{equation}
\frac{\tau (B^+)}{\tau (B_d)}\Bigg|_{\rm exp.}=
1.076(4)
\,  ,
\qquad
\frac{\tau (B_s)}{\tau (B_d)}\Bigg|_{\rm exp.}=
1.0032(32)
\, ,
\end{equation}
while the current state-of-the-art for their HQE predictions is summarised in Ref.~\cite{Albrecht:2024oyn} -- mostly based on Ref.~\cite{Lenz:2022rbq}, i.e.
\begin{equation}
\frac{\tau (B^+)}{\tau (B_d)}\Bigg|_{\rm HQE '22}=
1.086(22)
\,  ,
\qquad
\frac{\tau (B_s)}{\tau (B_d)}\Bigg|_{\rm HQE '22}=
\left\{
\begin{array}{cc}
  1.003(6)   &  {\rm Scenario \, B} \\[2mm]
  1.028(11)  &  {\rm Scenario \, A}
\end{array}
\right.
\, . 
\label{eq:tauBs/tauBd_HQE_old}
\end{equation}
The agreement between the HQE results and the experimental data is excellent for the $\tau(B^+)/\tau(B_d)$ lifetime ratio. As for $\tau(B_s)/\tau(B_d)$, the theory prediction is very sensitive to the value of the non-perturbative inputs that parametrise the two-quark operator matrix elements, particularly that of the Darwin operator due to its large short-distance coefficient~\cite{Lenz:2020oce, Mannel:2020fts}. This is reflected in the two different results shown in Eq.~\eqref{eq:tauBs/tauBd_HQE_old}, corresponding to two sets of parameters -- see Ref.~\cite{Lenz:2022rbq} for details on their definition. Specifically, while in the Scenario B, the experimental data and the HQE result again agree perfectly, in the second case (Scenario A), a slight tension arises. In this regard, it is worth noting that with the updated analysis performed in the present work, this small tension will be downsized.

The theoretical investigations have so far mainly focused on lifetime ratios, as the precision achievable for the total decay rates 
was strongly limited by the large uncertainties due to the free $b$-quark decay, which, until recently, was only known at the NLO-QCD accuracy~\cite{Bagan:1994zd,Bagan:1995yf,Lenz:1997aa,Lenz:1998qp,Krinner:2013cja,Greub:2000an,Greub:2000sy}.
Currently, the HQE predictions, based on NLO-QCD expressions for $\Gamma_b$, show in fact large uncertainties, in particular due to the sizeable renormalisation scale dependence~\cite{Lenz:2022rbq}. As the dominant NNLO-QCD corrections to non-leptonic $b$-quark decays have been recently determined in Ref.~\cite{Egner:2024azu}, 
we are now, for the first time, in the position to perform a comprehensive analysis at NNLO-QCD of both the free $b$-quark decay and the total decay rates of the $B_q$ mesons, and obtain more precise and stable theoretical predictions. This constitutes the main scope of this paper. 
Specifically, our study contains the following improvements with respect to previous analyses:
\begin{itemize}
    \item[$\diamond$] 
    We include NNLO-QCD corrections \cite{Egner:2024azu} to the free $b$-quark decay due to the current-current operators $Q_{1,2}$ in the $\Delta B = 1$ effective Hamiltonian, see Section~\ref{subsec:Heff}.
    Note that these do not yet lead to the complete NNLO result for $\Gamma_b$, since the corresponding NNLO-QCD corrections induced by the interference of the current-current and penguin operators $Q_{3,\ldots, 6}$, the insertion of the current-current operators into penguin diagrams of the effective theory, and the contribution of the chomomagnetic operator $Q_8$ 
    are still missing. 
    These corrections, however, are expected to yield a subleading effect. 
    
    \item[$\diamond$] 
    The NLO-QCD corrections to the chromo-magnetic operator in the $\Delta B = 0$ effective theory due to the non-leptonic decay $b \to c \bar{u} d$ have been determined very recently in Ref.~\cite{Mannel:2024uar}. At leading order, severe cancellations arise in the corresponding combination of $\Delta B = 1 $ Wilson coefficients, making the contribution from this operator particularly small at this order. This suppression, however, appears to be lifted once $\alpha_s$-corrections are included, and the NLO-QCD contributions to the chromo-magnetic operator have been found to lead to a sizeable shift of $1.9 \%$ in the non-leptonic width $\Gamma(b\to c \bar u d)$, whereas the effect amounts to only $-0.3 \%$ at LO~\cite{Mannel:2024uar}. Since the $\alpha_s$-corrections are not yet known for the decay $b \to c \bar{c} s$, a complete analysis at NLO-QCD is not yet possible, however we provide an estimate of the impact that these corrections might have on the total $B_q$-decay rates.

    \item[$\diamond$] A computation of the dimension-six Bag parameters for the $B_d$ meson entering the HQE prediction for $\tau(B^+)/\tau(B_d)$
    within the SM and beyond was very recently performed in Ref.~\cite{Bag-parameters-new} using the framework of the heavy-quark effective theory (HQET) sum rules. Although the authors confirm most of the results of the previous determinations~\cite{Kirk:2017juj, King:2021jsq}, the value of $\tilde B_3^d$, cf.~\eqref{eq:ME-dim-6-HQET-q-q}, turns out to be different. We implement the new results obtained in Ref.~\cite{Bag-parameters-new}, which lead, in particular, to a visible shift of the HQE prediction for $\tau(B^+)/\tau(B_d)$ from the current value~\cite{Lenz:2022rbq}.

    \item[$\diamond$] 
    We present a detailed comparison of different choices for the renormalisation scheme of both the bottom- and the charm-quark masses. Moreover, we also improve the analysis of the SU(3)$_F$ breaking effects in the non-perturbative parameter $\rho_D^3$ in the kinetic scheme, which leads, in particular, to changes in the lifetime ratio $\tau (B_s)/\tau (B_d)$ as compared to the previous work~\cite{Lenz:2022rbq}.

    \item[$\diamond$] 
    Using the state-of-the-art results for the HQE of semileptonic $B_q$ decays we provide theoretical predictions for the inclusive semileptonic branching fractions of the $B^+,B_d,$ and $B_s$ mesons, obtained entirely within the HQE. In these ratios, the theoretical uncertainties due to e.g.\ the CKM matrix element $V_{cb}$ and the fifth power of the $b$-quark mass, cancel, leading to a reduction of the theory error.

    \item[$\diamond$]
    Our final theory predictions for the total rates are performed at NNLO-QCD, that is using the same accuracy for the semileptonic modes as the one currently available for the non-leptonic ones. In particular the N$^3$LO-QCD corrections to the semileptonic $b$-quark decays computed in Ref.~\cite{Fael:2020tow} are not included. However, for completeness, the effect of adding all available corrections for the semileptonic channels, including QED and QCD corrections, on our results is also discussed.  

\end{itemize}
The paper is structured as follows: in Section~\ref{sec:Theory} we present the theoretical framework. Specifically we start in Section~\ref{subsec:Heff} describing the effective Hamiltonian for $\Delta B = 1$ transitions, in Section \ref{subsec:power} we summarize the status of power corrections within the HQE, and in Section \ref{subsec:mass} we discuss different choices for the renormalization schemes of the quark masses.
In Section~\ref{sec:numerics} we present our numerical analysis, starting from the description of the input parameters in Section~\ref{subsec:input} and followed by the discussion of our results. Specifically, in Section~\ref{subsec:free} we investigate the impact that different choices of the quark-mass schemes have on the value of the free $b$-quark decay at NNLO-QCD. In Section~\ref{subsec:total-widths} we show our predictions for the total widths of the $B^+, B_d$, and $B_s$ mesons in our default scenario, that is using the kinetic scheme for the bottom quark and the $\overline{\rm MS}$ scheme for the charm quark. Furthermore, we also present updated predictions for the corresponding lifetime ratios. Our results for the semileptonic branching ratios are discussed in Section~\ref{subsec:sl-Brs}. Finally, we conclude in Section~\ref{sec:conclusion}.

\section{Theoretical framework}
\label{sec:Theory}
\subsection{The heavy quark expansion}
\label{subsec:Heff}
Using the optical theorem, the total decay width of the $B_q$ meson $\Gamma(B_q)$ can be computed as
\begin{equation}
\Gamma (B_q) =    \frac{1}{2 m_{B_q}} {\rm Im}
\langle B_q | {\cal T}| B_q \rangle \, ,
\label{eq:Gamma_opt_th}
\end{equation}
with the transition operator given by
\begin{equation}
{\cal T}  =  
i \int d^4x 
\,  T \left\{ {\cal H} _{\rm eff} (x) \, ,
 {\cal H} _{\rm eff} (0)  \right\} \, .
 \label{eq:optical_theorem}
\end{equation}
The effective Hamiltonian ${\cal H}_{\rm eff}$ describes the weak decays of the $b$ quark, see e.g. the~review~\cite{Buchalla:1995vs}, and can be schematically decomposed as:\footnote{Note that we do not include the semileptonic operators relevant for the study of rare decays like $B \to K^{(*)} \gamma$, as the corresponding branching fractions are far below the current theoretical precision for the lifetimes.}
\begin{equation}
  {\cal H}_{\rm eff}  =   {\cal H}_{\rm eff}^{\rm NL} + {\cal H}_{\rm eff}^{\rm SL} 
  \,.
\label{eq:Heff-complete}
\end{equation}
The first term ${\cal H}_{\rm eff}^{\rm NL}$ parametrises the contribution due to non-leptonic $b$-quark transitions, i.e.
\begin{align}
  {\cal H}_{\rm eff}^{\rm NL} = 
  \frac{G_F}{\sqrt{2}} \sum_{q_3 = d, s}
  \left[\,
   \sum_{\substack{q_{1,2} = u, c} } \!\! \lambda_{q_1 q_2 q_3} 
  \Bigl(C_1 (\mu_b) \, Q_1^{q_1 q_2 q_3}  + C_2 (\mu_b) \, Q_2^{q_1 q_2 q_3}  \Bigr)
    -  \lambda_{q_3} 
  \!\! \!\! \sum \limits_{j=3, \ldots, 6, 8} \! \!\! C_j (\mu_b) \, Q_j^{q_3} 
   \right] + {\rm h.c.}\, ,
   \label{eq:Heff-NL}
\end{align}
where $\lambda_{q_1 q_2 q_3} = V_{q_1 b}^* V_{q_2 q_3} $
and $\lambda_{q_3} = V_{tb}^* V_{tq_3} $ denote the corresponding CKM factors, $C_i (\mu_b)$ are the Wilson coefficients of the $\Delta B = 1$ effective operators determined at the renormalisation scale $\mu_b \sim m_b$, and
$Q_{1,2}^{q_1 q_2 q_3}$, $Q_j^{q_3}$ with $j = 3, \ldots, 6,$ and $Q_8^q$, indicate respectively the current-current, the penguin and the chromo-magnetic operators. These are defined as following 
\begin{equation}
Q_1^{q_1 q_2 q_3} 
 =   
\left(\bar b^i \, \Gamma_\mu \, q_1^j \right)
\left(\bar q_2^j \, \Gamma^\mu  \, q_3^i \right)\,,
\qquad 
Q_2^{q_1 q_2 q_3} 
 =  
\left(\bar b^i  \, \Gamma_\mu  \, q_1^i \right)
\left(\bar{q}_2^j \, \Gamma^\mu  \, q_3^j \right)\,,
\label{eq:Q12}
\end{equation}
\begin{align}
Q_3^{q_3} 
& 
= (\bar b^i \, \Gamma_\mu \, q_3^i) \sum_{q} ( \bar q^j \, \Gamma^\mu \, q^j)
\,, \qquad 
Q_4^{q_3} = (\bar b^i \, \Gamma_\mu \, q_3^j) \sum_{q} (\bar q^j \, \Gamma^\mu \, q^i)\,, 
\label{eq:Q34}
\\
Q_5^{q_3} 
& 
=  (\bar b^i \, \Gamma_\mu \, q_3^i) \sum_{q} (\bar q^j \, \Gamma_+^\mu \, q^j)\,, 
\qquad
Q_6^{q_3} = (\bar b^i \, \Gamma_\mu \, q_3^j) \sum_{q} 
(\bar q^j \, \Gamma_+^\mu \, q^i)\,,
\label{eq:Q56}
\end{align}
\begin{equation}
Q_8^{q_3} = \frac{g_s}{8 \pi^2} m_b
\left(\bar b^i \, \sigma^{\mu\nu} (1 - \gamma_5) t^a_{ij} \, q_3^j \right) G^a_{\mu \nu}\,,
\label{eq:Q8}
\end{equation}
with $\Gamma_\mu = \gamma_\mu(1-\gamma_5)$,  $\Gamma_+^\mu = \gamma^\mu(1+\gamma_5)$ and $\sigma_{\mu \nu} =(i/2) [\gamma_\mu, \gamma_\nu]$. Moreover, in the above equations, $i,j = 1, 2, 3,$ and $a = 1, \ldots, 8,$ label the SU(3)$_c$ indices for fields respectively in the fundamental and in the adjoint representation, while in Eq.~\eqref{eq:Q8}, $g_s$~denotes the strong coupling, and $G_{\mu\nu} = G^a_{\mu\nu}t^a $ the gluon field strength tensor with $t^a_{ij}$ being the  SU(3)$_c$ generators. 

The Wilson coefficients $C_i$ are known at the NNLO-QCD accuracy~\cite{Gorbahn:2004my, Egner:2024azu}. Note that, as it is discussed in Ref.~\cite{Egner:2024azu}, because of the different convention adopted for the definition of the evanescent operators entering the NNLO computation of the leading power contribution, the results for $C_{1,2}$ given in Ref.~\cite{Egner:2024azu} differ from those of Ref.~\cite{Gorbahn:2004my}, though they coincide at NLO. 
A summary of the values of the Wilson coefficients for different choices of the scale $\mu_b$ and up to NNLO for $C_{1,2}$ \cite{Egner:2024azu}, NLO for $C_{3-6}$ \cite{Gorbahn:2004my}, and LO for $C_8^{\rm eff}$ \cite{Buchalla:1995vs} is presented in Table~\ref{tab:WCs}.

The second term in Eq.~\eqref{eq:Heff-complete} describes the semileptonic $b$-quark decays:
\begin{equation}
{\cal H}_{\rm eff}^{\rm SL} 
= 
\frac{G_F}{\sqrt 2} \sum_{q = u, c \,} \sum_{\, \ell = e, \mu, \tau}
V_{qb}^* \, Q^{q \ell} + {\rm h.c.}\,,
\label{eq:Heff-SL}
\end{equation}
with the corresponding semileptonic operator
\begin{equation}
    Q^{q \ell} =\left(\bar{b}\, \Gamma^\mu \, q \right)
\left(\bar \nu_\ell \, \Gamma_\mu \, \ell \right)\,.
\end{equation}

\begin{table}[t]
\renewcommand{\arraystretch}{1.3}
\centering
{
   \begin{tabular}{|C{1.8cm}|C{1.8cm}|C{1.8cm}|C{1.8cm}|C{1.8cm}|C{1.8cm}|C{1.8cm}|C{1.8cm}|}
   \hline 
    \multicolumn{2}{|c|}{$\mu _b\text{[GeV]}$} 
    & 2.5 & 4.2 & 4.5 & 4.8 & 9 \\
    \hline
    \multirow{3}{*}{$C_1 (\mu_b)$} 
    & NNLO & $-0.245$ & $-0.178$ & $-0.170$ & $-0.163$ & $-0.099$ \\
    & NLO & $-0.263$ & $-0.190$ & $-0.181$ & $-0.173$ & $-0.106$ \\
    & LO & $-0.351$ & $-0.267$ & $-0.257$ & $-0.249$ & $-0.174$ \\
 \hline
 \multirow{3}{*}{$C_2 (\mu_b)$} 
 & NNLO & $1.081$ & $1.055$ & $1.051$ & $1.049$ & $1.025$ \\
 & NLO  & $1.120$ & $1.081$ & $1.077$ & $1.073$ & $1.042$ \\
 & LO  & $1.163$ & $1.117$ & $1.112$ & $1.107$ & $1.070$ \\
 \hline
 \multirow{2}{*}{$C_3 (\mu_b)$} 
 & NLO & $0.019$ & $0.014$ & $0.013$ & $0.013$ & $0.008$ \\
 & LO & $0.016$ & $0.012$ & $0.012$ & $0.011$ & $0.007$ \\
 \hline
 \multirow{2}{*}{$C_4 (\mu_b)$}
 & NLO & $-0.046$ & $-0.036$ & $-0.035$ & $-0.033$ & $-0.024$ \\
 & LO & $-0.035$ & $-0.027$ & $-0.026$ & $-0.026$ & $-0.018$ \\
 \hline
 \multirow{2}{*}{$C_5 (\mu_b)$} 
 & NLO & $0.010$ & $0.009$ & $0.008$ & $0.008$ & $0.006$ \\
 & LO & $0.010$ & $0.008$ & $0.008$ & $0.007$ & $0.005$ \\
 \hline
 \multirow{2}{*}{$C_6 (\mu_b)$}
 & NLO & $-0.058$ & $-0.042$ & $-0.040$ & $-0.039$ & $-0.026$ \\
 & LO & $-0.047$ & $-0.034$ & $-0.033$ & $-0.031$ & $-0.021$ \\
 \hline
 $C_8^{\rm eff} (\mu_b)$ 
 & LO & $-0.165$ & $-0.152$ & $-0.150$ & $-0.149$ & $-0.136$ \\
 \hline
\end{tabular} 
}
\caption{Values of the $\Delta B=1$ Wilson coefficients for different choices of $\mu_b$.
For the input parameters we refer to Section~\ref{subsec:input}.}
\label{tab:WCs}
\end{table} 

In the framework of the HQE, the non-local operator in Eq.~\eqref{eq:optical_theorem} is evaluated by exploiting the fact that the $b$ quark is heavy i.e.\ $m_b \gg \Lambda_{\rm QCD}$, the latter defining a typical non-perturbative scale of the order of few hundreds MeV. Its momentum is then decomposed as following 
\begin{equation}
p_b^\mu = m_b  v^\mu + k^\mu\,,
\label{eq:c-quark-momentum}
\end{equation}
where $v^\mu = p_B^\mu/m_{B_q}$ is the four-velocity of the $B_q$ meson, and $k^\mu$ denotes a small residual momentum accounting for non-perturbative interactions of the $b$ quark with the light degrees of freedom inside the hadronic state, that is $k \sim \Lambda_{\rm QCD}$. 
The $b$-quark field is parametrised as  
\begin{equation}
b (x) = e^{ - i m_b v \cdot x} b_v (x)\,,  
\label{eq:phase-redef}    
\end{equation}
by factoring out the large component of the momentum and by introducing a rescaled field $b_v(x)$ containing only low oscillation  frequencies of the order of $k$. 
The field $b_v(x)$ is related to the HQET field $h_v(x)$, see e.g.\ the review~\cite{Neubert:1993mb}, by 
\begin{equation}
b_v (x) = h_v (x) + \frac{i \slashed D_\perp}{2 m_b} h_v (x)  
+ {\cal O} \left(\frac{1}{m_b^2} \right),
\label{eq:bv-hv-relation}
\end{equation}
with $D_\perp^\mu = D^\mu - (v \cdot D) \, v^\mu$, and the covariant derivative $D_\mu = \partial_\mu - i g_s A_\mu^a \, t^a $.
As result, within the HQE, the total decay width of the $B_q$ meson can be expressed in terms of the following operator product expansion (OPE)
\begin{equation}
\Gamma(B_q) = 
\Gamma_3  +
\Gamma_5 \frac{\langle {\cal O}_5 \rangle}{m_b^2} + 
\Gamma_6 \frac{\langle {\cal O}_6 \rangle}{m_b^3} + \ldots  
 + 16 \pi^2 
\left( 
  \tilde{\Gamma}_6 \frac{\langle \tilde{\mathcal{O}}_6 \rangle}{m_b^3} 
+ \tilde{\Gamma}_7 \frac{\langle \tilde{\mathcal{O}}_7 \rangle}{m_b^4} + \ldots 
\right),
\label{eq:HQE}
\end{equation}
where $\Gamma_d$ are short-distance coefficients that can be computed perturbatively in QCD, i.e.
\begin{equation}
\Gamma_d = \Gamma_d^{(0)} + \frac{\alpha_s}{\pi} \Gamma_d^{(1)} 
+ \left(\frac{\alpha_s}{\pi}\right)^2 \Gamma_d^{(2)} + \left(\frac{\alpha_s}{\pi}\right)^3 \Gamma_d^{(3)} + \ldots \, ,  
\label{eq:Gamma-i-pert-series}
\end{equation}
and $\langle {\cal O}_d \rangle \equiv
\langle B_q | {\cal O}_d |B_q  \rangle/(2 m_{B_q})$ denote the matrix element of the operators 
${\cal O}_d$, with mass dimension~$d$, of the $\Delta B = 0$ effective theory.
The leading term $\Gamma_3 = \Gamma_b$ describes the free $b$-quark decay. 
First power-suppressed corrections arise at order $1/m_b^2$ due to the kinetic and chromo-magnetic operators, schematically indicated by ${\cal O}_5$.
At order $1/m_b^3$ both contributions due to two-quark operators, e.g.~the Darwin operator, denoted by ${\cal O}_6$,
and four-quark operators, denoted by ${\tilde{\cal O}}_6$,
appear. Note that the latter originate from loop-enhanced diagrams, reflecting the explicit factor of $16 \pi^2$ in Eq.~\eqref{eq:HQE}.
The current status of the short-distance coefficients $\Gamma_d$ is summarised in Section~\ref{subsec:power}, for more details see also e.g.~Refs.~\cite{Albrecht:2024oyn, Lenz:2022rbq}.

\subsection{Status of power corrections}
\label{subsec:power}

In this section we provide a brief outline of the status of the HQE,
discussing the available results for the short-distance coefficients and the parametrisation of the hadronic matrix elements that enter our predictions for the $B_q$-mesons total decay widths. We start from the perturbative contributions.

For the free semileptonic $b$-quark decay, in the case of massless leptons, that is for $b\to c \ell^- \bar \nu_\ell$, with $\ell = e, \mu$, the N$^3$LO-QCD corrections have been determined in Ref.~\cite{Fael:2020tow}. For the tau-lepton mode, the result at NNLO-QCD can be derived from Ref.~\cite{Fael:2024gyw}, where the inclusive semileptonic differential width $d \Gamma_{\rm sl}/d q^2$ for a massive final-state lepton has been computed. 
As for the non-leptonic modes, as already  mentioned above, the accuracy reaches NNLO and the $\alpha_s^2$-corrections were recently determined in Ref.~\cite{Egner:2024azu}. 

At order $1/m_b^2$, the complete short-distance coefficient of the chromo-magnetic operator due to non-leptonic $b$-quark decays is known at the LO-QCD accuracy and can be found e.g.~in the Appendix of Ref.~\cite{Lenz:2020oce} --  originally determined in Refs.~\cite{Blok:1992he, Blok:1992hw, Bigi:1992ne}. Partial NLO-QCD corrections are also known. Very recently they have been computed for the case of one massive final state, namely for the $b \to c \bar u d$ transition~\cite{Mannel:2024uar, Mannel:2023zei}. The corresponding $\alpha_s$-corrections for the $b \to c \bar c s$ mode are still missing, hence a complete determination of $\Gamma_5^{(1)}$ is currently not yet possible. For the semileptonic case, the LO-QCD result is presented e.g.~in the Appendix of Ref.~\cite{Mannel:2017jfk}, first computed in Refs.~\cite{Balk:1993sz, Falk:1994gw}, while the NLO-QCD corrections have been determined in Refs.~\cite{Alberti:2013kxa, Mannel:2014xza, Mannel:2015jka} -- see also the recent works~\cite{Mannel:2021zzr, Moreno:2022goo}. 

At order $1/m_b^3$, both two- and four-quark operators contribute. For the former, in the nonleptonic case the current accuracy is limited to LO-QCD only, see Refs.~\cite{Lenz:2020oce, Mannel:2020fts, Moreno:2020rmk}. 
For the semileptonic decays, the coefficient of the Darwin operator has been first computed in Ref.~\cite{Gremm:1996df},
the generalisation to the case of two different final state masses is presented e.g. in Refs.~\cite{Rahimi:2022vlv, Moreno:2022goo}, while the corresponding NLO-QCD corrections have been determined in Refs.~\cite{Mannel:2019qel, Mannel:2021zzr, Moreno:2022goo}. As for the contribution of four-quark operators, the complete expressions for the dimension-six short-distance coefficients up to NLO-QCD corrections have been
obtained in Ref.~\cite{Franco:2002fc}, in the case of four-quark operators defined in HQET, and in Refs.~\cite{Franco:2002fc,Beneke:2002rj} for QCD operators. Note that for semileptonic modes the QCD corrections have been determined in Ref.~\cite{Lenz:2013aua}.

Finally, at order $1/m_b^4$, only the LO-QCD short-distance coefficients 
of the four-quark operators are known in the literature for both semileptonic and non-leptonic final states, see Refs.~\cite{Gabbiani:2003pq, Gabbiani:2004tp, Lenz:2013aua}. On the other hand, for semileptonic $b$-quark decays, power corrections up to $1/m_b^5$ have been computed, see e.g.\ Refs.~\cite{Dassinger:2006md,Mannel:2010wj,Mannel:2023yqf}.

We now turn to discuss the corresponding hadronic matrix elements. We define the dimension-six four-quark operators as 
\begin{align}
		{\tilde O}_1^q  
		& =  
		(\bar{h}_v\,\gamma_\mu (1-\gamma_5) q)\,(\bar{q}\,\gamma^\mu (1-\gamma_5) h_v) ,
		\label{eq:O1-HQET} \\[1mm]
		{\tilde O}_2^q  
		& =  
		(\bar{h}_v (1 - \gamma_5) q)\,(\bar{q} (1 + \gamma_5) h_v) ,
		\label{eq:O2-HQET} \\[1mm]
		{\tilde O}_3^q  
		& =  
		(\bar{h}_v \, \gamma_\mu (1-\gamma_5) \, t^a q) 
		\, (\bar{q} \, \gamma^\mu (1-\gamma_5) \, t^a  h_v) ,
		\label{eq:T1-HQET} \\[1mm]
		{\tilde O}_4^q 
		& =  
		(\bar{h}_v (1-\gamma_5) t^a q)\,(\bar{q}(1 + \gamma_5) t^a h_v),
		\label{eq:T2-HQET}
\end{align}
where $h_v$ is the HQET field. Their matrix elements are parametrised as
\cite{King:2021jsq}
	\begin{eqnarray}
		\langle {B}_q | {\tilde O}_i^q \, | {B}_q \rangle 
		& = & 
		F_q^2(\mu_0) \, m_{B_q} \, \tilde B_i^q(\mu_0) \,,
		\label{eq:ME-dim-6-HQET-q-q}
		\\[2mm]
		\langle {B}_q | {\tilde O}_i^{q^\prime} | {B}_q \rangle 
		& = & 
		F_q^2 (\mu_0) \, m_{B_q} \, \tilde \delta^{q^\prime q}_i (\mu_0)\,,
		\qquad q \not = q^\prime \,,
		\label{eq:ME-dim-6-HQET-q-q-prime}
	\end{eqnarray}
where $\tilde B_i^q(\mu_0)$ and $\tilde \delta_i^{q^\prime q}(\mu_0)$, are the corresponding Bag parameters and ``eye-contractions'' evaluated at $\mu_0$, the renormalisation scale of the $\Delta B =0$ operators, and $F_q(\mu_0)$ is the HQET decay constant. Note that by means of the relation~\cite{Neubert:1992fk}
\begin{equation}
		f_{B_q} = \frac{F_q (\mu_0)}{\sqrt{m_{B_q}}} \left[1 +  
		\frac{\alpha_s(\mu_0)}{2\pi} 
		\left(\ln \left(\frac{m_b^2}{\mu_0^2} \right)
		- \frac 4 3 \right) + 
		{\cal O}\left(\frac{1}{m_b}\right) \right] \,,
		\label{eq:decay-const-conv}
	\end{equation}	
we can express our results in terms of the QCD decay constant $f_{B_q}$ which is determined very precisely from lattice QCD.

At order $1/m_b^4$ the number of independent operators largely increases and for simplicity we do not show them explicitly here but refer to Ref.~\cite{King:2021xqp} for their expressions. 

Finally, in the case of two-quark operators at order $1/m_b^2$ and $1/m_b^3$, the matrix elements of the kinetic, chromo-magnetic and Darwin operators are parametrised in terms of the inputs $\mupi$, $\muG$, $\rhoD$ as
\begin{align}
		2 m_{B_q} \, \mu_\pi^2 (B_q) 
		& =  
		- \langle B_q |\bar{b}_v (i D_\mu)(i D^\mu) b_v | B_q \rangle  \, ,
		\label{eq:mupi-def}
		\\
		2 m_{B_q} \, \mu_G^2 (B_q) 
		& =  
		\langle B_q | \bar{b}_v (i D_\mu)(i D_\nu) (-i \sigma^{\mu \nu}) b_v | B_q \rangle 
		\, , 
		\label{eq:muG-def}
		\\
		2 m_{B_q} \, \rho_D^3 (B_q) & =  
		\langle B_q | \bar{b}_v (i D_\mu)(i v \cdot D) (i D^\mu) b_v | B_q \rangle
		\, .
		\label{eq:rhoD-def}
	\end{align}

\subsection{Renormalisation schemes for the quark masses}
\label{subsec:mass}
For the numerical evaluation of the decay rates we use different 
renormalisation schemes for the quark masses. This is particularly 
relevant for $\Gamma_3$ where NNLO-QCD corrections are now available, but it
also affects the power-suppressed terms.

The starting point is the expression for $\Gamma_3$ in which both the charm and bottom
quark masses are renormalised on-shell (OS). The NLO and NNLO calculations
are most conveniently performed in this scheme. The corresponding numerical
results will be discussed in Section~\ref{subsub::pole}.

For the conversion of the pole masses to the $\overline{\rm MS}$
scheme we use two-loop relations~\cite{Gray:1990yh} including
corrections from closed massless and massive quark loops which appear
for the first time at order $\alpha_s^2$~\cite{Gray:1990yh} (see
Refs.~\cite{Bekavac:2007tk, Fael:2020bgs} for convenient analytical
expressions). For both the charm and the bottom quark relations we
assume that the up, down and strange quarks are massless and take into
account the exact mass dependence on the heavy quarks. Top quark
effects are suppressed and are not considered. Our final results for the decay rates are expressed in terms of $\overline{m}_b^{(5)}(\mu_b)$ and
$\overline{m}_c^{(4)}(\mu_c)$ where the superscript indicates the number of
active quark flavours. In our phenomenological analysis we keep $\mu_b$ and $\mu_c$ separate.
Results for the partonic decay rate in the 
$\overline{\rm MS}$ scheme are discussed in Section~\ref{subsub::MSbar}.

In our default choice for the renormalization scheme, we transform the charm-quark pole mass
in the $\overline{\rm MS}$ scheme, just as described above, while the bottom-quark pole mass is converted to the kinetic scheme using the following relation, valid up to ${\cal O}((1/m_b^{\rm kin})^2)$ corrections~\cite{Bigi:1996si}
\begin{eqnarray}
  m_b^{\rm OS} \equiv m_b^{\rm kin} (0) &=& m_b^{\rm kin} (\mu^{\rm
    cut}) + [\Lambda (\mu^{\rm cut})]_{\rm pert} + \frac{[\mu_\pi^2
      (\mu^{\rm cut})]_{\rm pert}}{2 m_b^{\rm kin} (\mu^{\rm cut})}
  \,,
  \label{eq:mb-kin}
\end{eqnarray}
where the Wilsonian cutoff $\mu^{\rm cut}$ is taken to be of the order of $1~{\rm GeV}$. 
In the kinetic scheme, also the non-perturbative parameters $\mu_\pi^2$ and $\rho_D^3$ have to be transformed according to
\begin{eqnarray}
  \mu_\pi^2(0) &=& \mu_\pi^2 (\mu^{\rm cut}) - [\mu_\pi^2 (\mu^{\rm
      cut})]_{\rm pert}\,,
  \qquad 
  \label{eq:mupi-kin}\\ 
  \rho_D^3(0) &=& \rho_D^3 (\mu^{\rm cut}) - [\rho_D^3 (\mu^{\rm
      cut})]_{\rm pert}\,.
  \label{eq:rhoD-kin}
\end{eqnarray}
The perturbative quantities introduced in Eqs.~(\ref{eq:mb-kin}) -- (\ref{eq:rhoD-kin}) are known to three-loop
order~\cite{Fael:2020iea, Fael:2020njb}, however, in our analysis we
use for consistency the two-loop relation~\cite{Czarnecki:1997sz}.
 The explicit expressions for $[\Lambda (\mu^{\rm cut})]_{\rm pert}$, $[\mu_\pi^2 (\mu^{\rm
      cut})]_{\rm pert}$, $[\rho_D^3 (\mu^{\rm
      cut})]_{\rm pert}$ as function of the Wilsonian cutoff $\mu^{\rm cut}$ are given in the Appendix of Ref.~\cite{Fael:2020njb}. Our results for the partonic decay rate in this scenario are discussed in Section~\ref{sub:kin+MS}, while our final predictions for the total decays widths are presented in Section~\ref{subsec:total-widths}.

\section{Numerical results at NNLO-QCD}
\label{sec:numerics}

\subsection{Numerical values for the inputs}
\label{subsec:input}

\begin{table}[t]
\renewcommand{\arraystretch}{1.25}
\begin{center}
\begin{tabular}{|c|c||c|c|}
\hline
Parameter & Source & Parameter & Source \\
\hline
$m_Z = 91.1880 \GeV$ & \cite{PDG:2024} &
$m_W = 80.3692 \GeV$ & \cite{PDG:2024} \\
$m_t = 172.57 \GeV$  & \cite{PDG:2024} &
$\alpha_s^{(5)} (M_Z) = 0.1180 \pm 0.0010$ & \cite{PDG:2024}\\
\hline
\end{tabular}
\caption{\label{tab::input1}Input values for the SM parameters.}
\end{center}
\end{table}

\begin{table}[t]
\centering
\renewcommand{\arraystretch}{1.4}
    \begin{tabular}{|c||c|c||c|c|}
    \hline 
    Parameter 
    & $B^{+}, B_{d}$ 
    & Source
    & $B_s$
    & Source \\
    \hline
    \hline
    $\mu_\pi^2 (B_q)$ [GeV$^2$]
    & $0.454 \pm  0.043 $
    & Exp. fit \cite{Finauri:2023kte}
    & $ 0.534 \pm  0.074 $
    & Exp. fit + Eq.~\eqref{eq:SU3f-break-mupi} 
    \\
    \hline
    $\mu_G^2 (B_q)$ [GeV$^2$]
    & $0.274 \pm  0.053 $
    & Exp. fit \cite{Finauri:2023kte} 
    & $ 0.321 \pm  0.072$ 
    & Exp. fit + Eq.~\eqref{eq:SU3f-break-muG}
    \\
    \hline
    $\rho_D^3 (B_q)$ [GeV$^3$]
    & $0.176 \pm  0.019 $
    & Exp. fit \cite{Finauri:2023kte}
    & $0.210 \pm  0.034$ 
    & Exp. fit +  Eq.~\eqref{eq:SU3f-break-rhoD-EoM} \\
    \hline
    \end{tabular}
    \caption{Input values of the two-quark non-perturbative parameters used in our analysis, defined in the kinetic scheme and in correspondence of $\mu^{\rm cut} = 1$ GeV.
    }
    \label{tab:Non-perturbative-input}
\end{table}

\begin{table}[th]
\centering
\renewcommand{\arraystretch}{1.5} 
\begin{tabular}{|c||c|c|c|c|}
\hline
$\mu_0 = 1.5\GeV$    
& $ \tilde B_1^q$ 
& $ \tilde B_2^q$ 
& $ \tilde B_3^q$ 
& $ \tilde B_4^q$ 
\\
\hline
\hline
    $\langle B_{u,d} | \tilde O^{u,d}_i | B_{u,d} \rangle$ 
     & $1.0026^{+0.0246}_{-0.0221}$ 
     & $0.9982^{+0.0206}_{-0.0214}$ 
     & $-0.0057^{+0.0221}_{-0.0225}$ 
     & $-0.0014^{+0.0216}_{-0.0221}$
\\
\hline
     $\langle B_{s} | \tilde O^{s}_i | B_{s} \rangle$  
     & $1.0022^{+0.0246}_{-0.0221}$  
     & $0.9983^{+0.0246}_{-0.0221}$ 
     & $-0.0036^{+0.0265}_{-0.0270}$  
     & $-0.0009^{+0.0259}_{-0.0265}$
\\
\hline
\hline
$\mu_0 = 1.5 \GeV $    
& $ \tilde \delta^{q^\prime q}_1$
& $ \tilde \delta^{q^\prime q}_2$ 
& $ \tilde \delta^{q^\prime q}_3$ 
& $ \tilde \delta^{q^\prime q}_4$ 
\\
\hline
\hline
$\langle B_{d,u} | \tilde O^{u,d}_i | B_{d,u} \rangle $
& $0.0026^{+0.0142}_{-0.0092}$ 
& $-0.0018^{+0.0047}_{-0.0072}$ 
& $-0.0004^{+0.0015}_{-0.0024}$ 
& $0.0003^{+0.0012}_{-0.0008}$
\\
\hline
$ \langle B_s | \tilde O^{u,d}_i | B_s \rangle$ 
& $0.0025^{+0.0144}_{-0.0093}$ 
& $-0.0018^{+0.0047}_{-0.0072}$ 
& $-0.0004^{+0.0015}_{-0.0024}$ 
& $0.0003^{+0.0012}_{-0.0008}$
\\
\hline
$\langle B_{d,u} | \tilde O^s_i | B_{d,u} \rangle$ 
& $0.0023^{+0.0140}_{-0.0091}$ 
& $-0.0017^{+0.0046}_{-0.0070}$ 
& $-0.0004^{+0.0015}_{-0.0023}$ 
& $0.0003^{+0.0012}_{-0.0008}$
\\
\hline
\end{tabular}
\caption{Input values of the dimension-six Bag parameters employed in our analysis at the renormalisation scale $\mu_0 = 1.5\GeV$. See the text for details and references.
}
\label{tab:Bag-parameters}
\end{table}

In Table~\ref{tab::input1} we list the input values for the masses of the $W$ and $Z$
bosons and the top quark, which enter the calculation of the Wilson
coefficients, as well as of the strong coupling constant. For the predictions of
the decay rates we need $\alpha_s^{(n_f)}(\mu_s)$ at the scale $\mu_s\sim m_b$ with
four and five active flavours $n_f$. This is obtained using the five-loop running
and, for the case of $\alpha_s^{(4)}$, also the four-loop decoupling relations as
implemented in {\tt RunDec}~\cite{Herren:2017osy}. Note that we decouple the
bottom quark at twice the bottom quark mass and set $\mu_s = \mu_b$. A comparison of the values of the Wilson coefficients of the
$\Delta B=1$ theory at different renormalisation scales $\mu_b$ is shown in Table~\ref{tab:WCs}.

As previously stated, in our default scenario we renormalise the bottom quark mass in the kinetic scheme, while the charm quark mass is defined in the $\overline{\rm MS}$ scheme. For the input values of the bottom and charm quark masses we use the results from the recent fit to data on inclusive
semileptonic $B$ decays~\cite{Finauri:2023kte}, namely 
\begin{eqnarray}
  m_b^{\rm kin} (1\GeV) & = & (4.573 \pm 0.012) \GeV,
  \label{eq:mb-kin-SL-fit}
  \\
  \overline{m}_c (2\GeV) & = & (1.090 \pm 0.010) \GeV,
  \label{eq:mc-MS-2GeV-SL-fit}
\end{eqnarray}
in correspondence of the cutoff scale $\mu^{\rm cut}=1$~GeV
and of the charm renormalisation scale
$\mu_c=2$~GeV. 
Starting from $m_b^{\rm kin}$ in Eq.~(\ref{eq:mb-kin-SL-fit}), the bottom-quark mass in the $\overline{\rm MS}$ scheme is given to three-loop accuracy~\cite{Fael:2020iea} by 
\begin{eqnarray}
  \overline{m}_b (\overline{m}_b) &=& 4.216 \GeV.
  \label{eq:mb_MSbar}
\end{eqnarray}
For later convenience we also provide the values of the charm quark mass obtained from Eq.~(\ref{eq:mc-MS-2GeV-SL-fit}) and run to the scales $\mu_c=3$~GeV, $\overline{m}_b$ and $m_b^{\rm kin}$, respectively. We obtain
\begin{eqnarray}
  \overline{m}_c (3 \GeV) & = & 0.9847 \GeV  \, ,
  \label{eq:mc-MS-3GeV-SL-fit}
  \\
  \overline{m}_c (\overline{m}_b) & = & 0.9184  \GeV \, ,
  \label{eq:mc-MS-mbMS-SL-fit}
  \\
   \overline{m}_c (m_b^{\rm kin}) & = & 0.9046  \GeV \, .
  \label{eq:mc-MS-mbkin-SL-fit}
\end{eqnarray}

The non-perturbative parameters $\mupi$,
$\muG$, and $\rhoD$, are also converted in the kinetic scheme and, unless otherwise stated, they must be understood in this scheme. In the case of the $B_d$ and $B^+$ mesons, their values can be obtained from the recent fit to semileptonic $B$ decays~\cite{Finauri:2023kte}, see Table~\ref{tab:Non-perturbative-input}.\footnote{Note that our definition of the chromo-magnetic operator differs from the one of Ref.~\cite{Finauri:2023kte}, hence the different value of $\mu_G^2(B)$ in Table~\ref{tab:Non-perturbative-input}, as compared to the one given in Ref.~\cite{Finauri:2023kte}, i.e.\ $\mu_G^2=(0.288\pm0.049) \GeV^2$.} The corresponding inputs for the $B_s$ meson are mostly unknown,\footnote{First steps in determining these parameters from data on exclusive $B_s$ decays have been taken in Ref.~\cite{DeCian:2023ezb}.} however, they can be derived estimating the size of the SU(3)$_F$-breaking effects in the corresponding non-perturbative parameters as outlined below -- see also the work~\cite{Lenz:2022rbq} for more details.

In the literature there are several determinations of the SU(3)$_F$-breaking effects in the dimen\-sion-five non-perturbative parameters $\mu_\pi^2$ and $\mu_G^2$. They are
based on the use of spectroscopy relations, see e.g.\ Refs.~\cite{Uraltsev:2001ih, Bigi:2011gf}, and, more recently, on lattice QCD computations~\cite{Gambino:2017vkx, Gambino:2019vuo}.
In our analysis, we consider an interval for the size of these effects which conservatively covers all the available estimates, namely we use
\begin{equation}
\mupis - \mu_\pi^2(B) \approx  (0.08 \pm 0.06) \GeV^2 \,,
\label{eq:SU3f-break-mupi}
\end{equation}
and
\begin{equation}
\frac{\muGs}{\mu_G^2(B)} \simeq  1.17 \pm 0.13 \,,
\label{eq:SU3f-break-muG}
\end{equation}
where $B$ refers to either the $B_d$ or $B^+$ meson in the isospin limit. Combining the above results for the SU(3)$_F$-breaking effects with the values of $\mu_\pi^2(B)$ and $\mu_G^2(B)$ obtained from the semileptonic fit~\cite{Finauri:2023kte} we arrive at the estimates for $\mu_\pi^2(B_s)$ and $\mu_G^2(B_s)$ shown in Table~\ref{tab:Non-perturbative-input}.

As for the Darwin parameter $\rho_D^3$, the size of the SU(3)$_F$-breaking effects is obtained using the equations of motion (EOM) for the gluon field strength tensor, which allow to rewrite the matrix element of the Darwin operator in terms of the dimension-six four-quark operator matrix elements~\cite{Bigi:2011gf}. This yields~\cite{Lenz:2022rbq}
\begin{equation}
\left[\frac{\rhoDs}{\rho_D^3(B)}\right]^{\rm OS} \!\!\!
\approx \frac{f_{B_s}^2 \, m_{B_s}}{f_B^2 \, m_B}
\approx 1.49 \pm 0.25\,,
\label{eq:SU3f-break-rhoD-EoM}
\end{equation}
where we have used lattice results for the decay constants
\cite{Aoki:2019cca}, see Eq.~\eqref{eq:decay-constants},
and additionally assigned a conservative uncertainty of $50\%$
to the SU(3)$_F$-symmetric limit to account for missing power corrections.
The superscript in Eq.~\eqref{eq:SU3f-break-rhoD-EoM} has been added to explicitly indicate that the above relation refers to the matrix elements of the Darwin operator as defined in the OS scheme. Using relation~\eqref{eq:rhoD-kin} and taking into account that $[\rho_D^3(\mu^{\rm cut})]_{\rm pert}$ is independent of the specific $B$-meson state, 
we arrive at the following expression for $\rho_D^3 (B_s)$ in the kinetic scheme, namely
\begin{equation}
[\rhoDs]^{\rm kin} = 
\left([\rho_D^3(B)]^{\rm kin} - [\rho_D^3(\mu^{\rm cut})]_{\rm pert}\right) \left[\frac{\rhoDs}{\rho_D^3(B)}\right]^{\rm OS} \!\!\! + [\rho_D^3(\mu^{\rm cut})]_{\rm pert}\,.
\label{eq:rhoD-kin-Bs}
\end{equation}
The quantities $[\rho_D^3(B)]^{\rm kin}$ and $[\rho_D^3(\mu^{\rm cut})]_{\rm pert}$ depend on the Wilsonian cutoff $\mu^{\rm cut}$, and for $\mu^{\rm cut} = 1\GeV$, using the inputs listed in Section~\ref{subsec:input}, we obtain the following estimate for $\rho_D^3(B_s)$ in the kinetic scheme, corresponding to a violation of the SU(3)$_F$-symmetry of $\approx 19 \%$, namely
\begin{equation}
{\left[\rho_D^3 (B_s)\right]^{\rm kin}} \simeq (0.210 \pm 0.034) \GeV^3\,.
\end{equation}
Note that, compared with what was done in Ref.~\cite{Lenz:2022rbq}, in the present analysis we do not consider two different sets of values for the non-perturbative inputs of the two-quark operators. In the previous work, the need to distinguish between the two scenarios was strongly motivated by the fact that two different values for the parameter $\rho_D^3$ were available in the literature, as obtained from fits to data on inclusive semileptonic $B$-meson decays which used, on the one hand, the $q^2$-moments~\cite{Bernlochner:2022ucr}, and, on the other, the hadronic mass spectrum and lepton energy moments~\cite{Bordone:2021oof}. These studies have been superseded by the recent analysis performed in Ref.~\cite{Finauri:2023kte}, where all available data and also partial $\alpha_s^2$-corrections to the $q^2$ spectrum were included. Additionally, by taking into account the perturbative corrections to the parameter $\rho_D^3$, cf.~Eq.~\eqref{eq:rhoD-kin-Bs}, we now obtain a value for the SU(3)$_F$ breaking in this parameter which is significantly smaller than what was used in the Scenario A of Ref.~\cite{Lenz:2022rbq}. Hence, the set of inputs adopted in the present work falls between the two cases considered previously, making a distinction no longer necessary. 

For the matrix elements of the dimension-six four-quark operators, see Table~\ref{tab:Bag-parameters},
we use the updated results for the Bag parameters of the $B_d$ meson computed in Ref.~\cite{Bag-parameters-new}, while in the case of the $B_s$ meson
we combine the results from Ref.~\cite{Bag-parameters-new} with those of Ref.~\cite{King:2021jsq}, where the size of SU(3)$_F$ breaking effects in these parameters was determined. Notice that the values of $\tilde B_{1,2}^s$ are unchanged, but those of $\tilde B_{3,4}^s$ differ from the ones in Ref.~\cite{King:2021jsq}. 
In the latter case, we also include an additional $20\%$ uncertainty. As for the remaining parameters i.e.\ the so-called ``eye-contractions'' $\tilde \delta^{q^\prime q}_i$, their values are taken from Ref.~\cite{King:2021jsq} and note that the contributions with $q = q^\prime$ are actually included in the definition of the $\tilde B_i^q$.
Finally, we emphasise that the dimension-six Bag parameters given in Refs.~\cite{Bag-parameters-new, King:2021jsq} are shown at the renormalisation scale $\mu_0 = 1.5$ GeV, and that the values used in our analysis are obtained by evolving these parameters to the scale $\mu_0 = m_b^{\rm kin}$ using the one-loop running~\cite{King:2021jsq}. However, the running of the $\tilde \delta_i^{q\prime q}$ is neglected, as these parameters already represent corrections of ${\cal O}(\alpha_s)$.

At dimension-seven we employ vacuum insertion approximation. In this case, the parametrisation of the corresponding matrix elements depends on the quantity $\bar \Lambda_{(s)}$, see Ref.~\cite{Lenz:2022rbq}, defined as
$\bar \Lambda_q = m_{B_q}- m_b$, and we use, respectively~\cite{King:2021jsq}
\begin{equation}
   \bar \Lambda = (0.5 \pm 0.1) {\rm GeV}\,, 
   \qquad
   \bar \Lambda_s = (0.6 \pm 0.1) {\rm GeV}\,,
\end{equation}
where $\bar \Lambda$ refers to the parameter for either the $B_d$ or $B^+$ meson in the isospin limit.

As for the remaining parameters, the values of the decay constants are determined very precisely from lattice QCD and we use~\cite{Aoki:2019cca}
\begin{equation}
    f_B = (0.1900\pm 0.0013)\GeV \,,
    \qquad
    f_{B_s} = (0.2303 \pm 0.0013)\GeV\,.
    \label{eq:decay-constants}
\end{equation}
The values of the $B_q$-meson masses, known with very high precision, are taken from the PDG~\cite{PDG:2024}, that is
\begin{equation}
m_{B^+} = 5.27934 \GeV, \quad
m_{B_d} = 5.27965 \GeV, \quad
m_{B_s} = 5.36688 \GeV \, .
\end{equation}

Concerning the CKM matrix elements, we adopt the standard parametrisation in terms of $\theta_{12}, \theta_{13}, \theta_{22}, \delta$, which is then expressed in terms of the four independent parameters $|V_{us}|$, $|V_{cb}|$, $|V_{ub}/V_{cb}|$, and $\delta$. The value of $V_{cb}$ is taken from the latest semileptonic fit~\cite{Finauri:2023kte}, i.e.
\begin{equation}
|V_{cb}| = (41.97 \pm 0.48) \times 10^{-3},  
\label{eq:Vcb-value}
\end{equation}
while for the remaining CKM inputs we use the results of the global CKM fit quoted by the CKMfitter group~\cite{Charles:2004jd} (online update), namely
\begin{eqnarray}
|V_{us}| & = & 
0.22498^{+0.00023}_{-0.00022} \, , \\
\frac{|V_{ub}|}{|V_{cb}|} & = & 
0.08887^{+0.00141}_{-0.00154} \, ,
\\[1mm]
\delta & = & 
\left(66.23^{+0.60}_{-1.43}\right)^\circ\,. 
\end{eqnarray}
Finally, the theoretical uncertainties are obtained by varying all the input parameters within the corresponding uncertainty ranges, taking into account all correlations, when available. For the renormalisation scales, we vary 
$\mu_b$, $\mu_c$, and $\mu_0$ independently in the interval $m_b^{\rm kin}/2 \leq \mu_b,\mu_c,\mu_0 \leq 2 m_b^{\rm kin}$, and the Wilsonian cutoff $\mu^{\rm cut}$ in the range $0.7 \GeV \leq \mu^{\rm cut} \leq 1.3 \GeV$. In the latter case, the dependence on $\mu^{\rm cut}$ of the parameters $\rho_D^3$ and $\mu_\pi^2$ defined in the kinetic scheme is also included. The corresponding uncertainties due to the variation of all four scales are then added in quadrature.
Finally, the quoted errors in our results are obtained by adding in quadrature the total parametric uncertainty and the one due to the renormalisation scale variation. 

\subsection{Partonic decay at NNLO-QCD}
\label{subsec:free}
In this section we discuss our predictions, at NNLO-QCD, of the free $b$-quark decay contribution $\Gamma_3$ for different choices of the bottom- and charm-quark masses. 
The results at NNLO for the non-leptonic modes are taken from Ref.~\cite{Egner:2024azu}, while for the semileptonic modes they are obtained by integrating over $q^2$ the corresponding differential distribution $d \Gamma_{\rm sl}/d q^2$ computed in Ref.~\cite{Fael:2024gyw}.  
For the bottom- and charm-quark masses we consider, respectively, the pole scheme, 
the $\overline{\rm MS}$ scheme and the case where the bottom-quark mass is renormalised in the kinetic scheme and the charm quark in the $\overline{\rm MS}$ scheme. Note that in this case the perturbative contributions $[\mu_\pi^2(\mu^{\rm cut})]_{\rm pert}$ and $[\rho_D^3(\mu^{\rm cut})]_{\rm pert}$ in Eqs.~\eqref{eq:mupi-kin}, \eqref{eq:rhoD-kin} are also included as part of the NLO- and NNLO-QCD corrections to $\Gamma_3$.
The latter scheme also represents our default scenario.

\subsubsection{Pole scheme}
\label{subsub::pole}
\begin{figure}[t]
\centering
\includegraphics[scale=0.5]{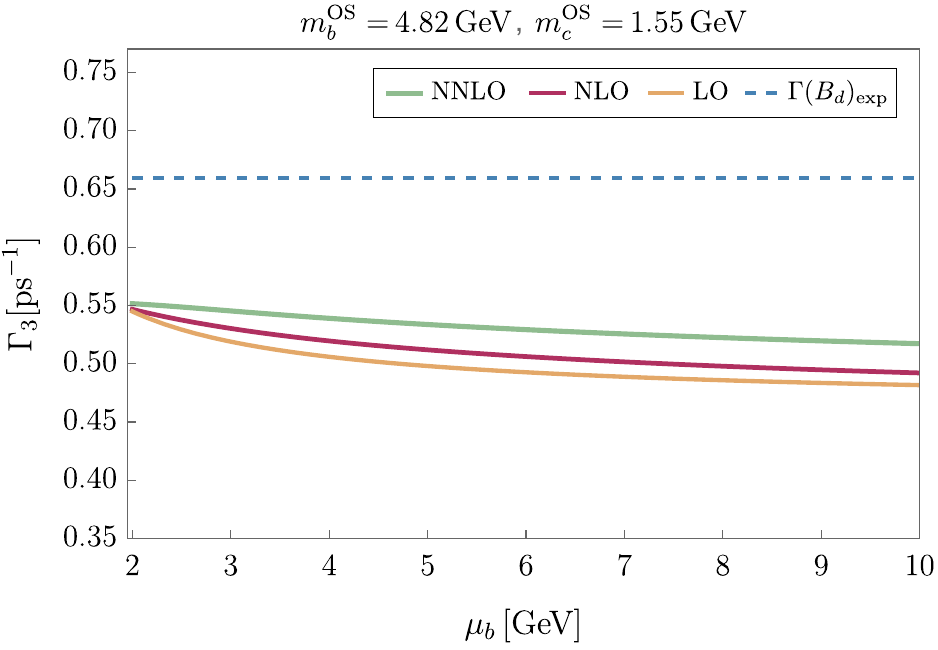} \,
\includegraphics[scale=0.5]{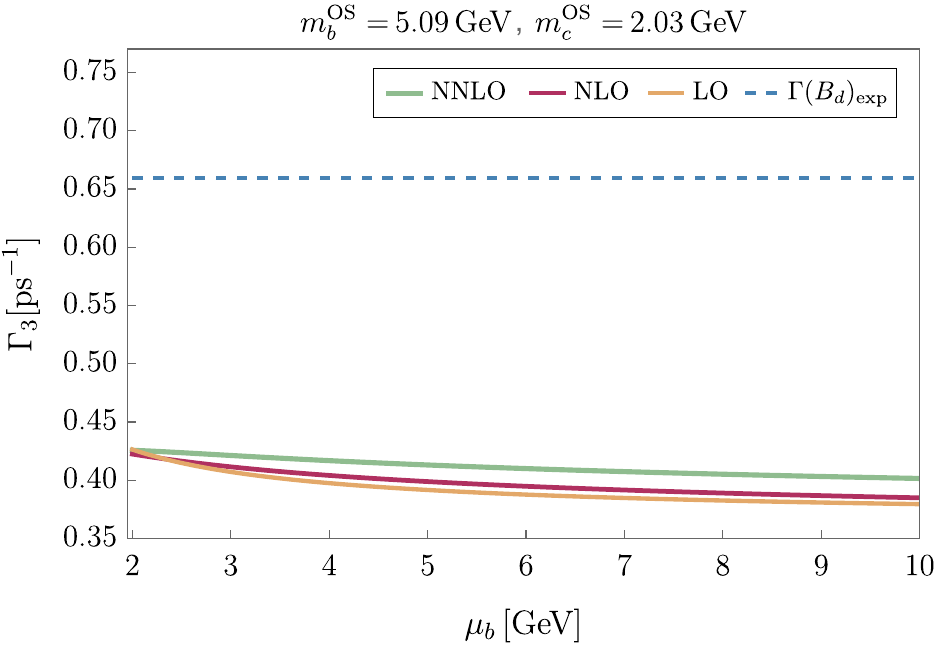}
\caption{Results for the free $b$-quark decay rate $\Gamma_3$ at NNLO- (solid green), NLO- (solid magenta), and LO-QCD (solid orange) in the OS scheme, using respectively two-loop (left) and four-loop (right) conversion relations from the $\overline{\rm MS}$ scheme. The experimental value of $\Gamma(B_d)$ (dashed blue) is also shown for an approximate reference.}
\label{fig:Gamma3-NNLO-vs-NLO-vs-LO_Pole}
\end{figure}

To obtain the values of the quark masses in the on-shell scheme,
we use as input the results for the bottom- and charm-quark masses in the $\overline{\rm MS}$ scheme given in Eqs.~\eqref{eq:mb_MSbar} and~\eqref{eq:mc-MS-3GeV-SL-fit}, and employ the corresponding conversion relations to the OS scheme as implemented in the package {\tt RunDec}~\cite{Herren:2017osy}. Using
two- and four-loops accuracy, we obtain
\begin{eqnarray}
m_b^{\rm OS} & = & 4.82 \GeV\,, \qquad m_c^{\rm OS} = 1.55 \GeV\,,
\qquad \mbox{(2 loops)}\,,
\label{eq:OS-masses-2L}
\\
m_b^{\rm OS} & = & 5.09 \GeV\,, \qquad m_c^{\rm OS} = 2.03 \GeV\,,
\qquad \mbox{(4 loops)}\,.
\label{eq:OS-masses-4L}
\end{eqnarray}
In this scenario, using the first set of inputs in Eq.~\eqref{eq:OS-masses-2L} our result for $\Gamma_3$ at $\mu_b = m_b^{\rm OS}$ reads
\begin{eqnarray}
\Gamma_3 & = & \Gamma_0 
\biggl[3.138 + 1.279 \, \frac{\alpha_s}{\pi}  
+ 28.83 \left(\frac{\alpha_s}{\pi}\right)^{\!2\,}
\biggr],     
\label{eq:Gamma3-decomp_Pole_A}
\end{eqnarray}
where $\Gamma_0 = 0.1591 \psinv$, see the definition in Eq.~\eqref{eq:Gamma_0}, and  $\alpha_s \equiv \alpha_s^{(4)}(\mu_b)$. On the other hand, using the second set of inputs given in Eq.~\eqref{eq:OS-masses-4L} we obtain
\begin{eqnarray}
\Gamma_3 & = & \Gamma_0 
\biggl[1.873 + 0.5136 \, \frac{\alpha_s}{\pi}  
+ 15.24 \left(\frac{\alpha_s}{\pi}\right)^{\!2\,}
\biggr],     
\label{eq:Gamma3-decomp_Pole_B}
\end{eqnarray}
with $\Gamma_0 = 0.2089 \psinv$. The renormalisation scale dependence of $\Gamma_3$ in Eqs.~\eqref{eq:Gamma3-decomp_Pole_A} and \eqref{eq:Gamma3-decomp_Pole_B} is shown, respectively, on the left and right panels of Fig.~\ref{fig:Gamma3-NNLO-vs-NLO-vs-LO_Pole}, together with the experimental result for $\Gamma(B_d)$, which we use as an approximate value for $\Gamma_3$.\footnote{Note that in the case of the $B^+$ meson, a large Pauli interference contribution from four-quark operators gives a sizeable correction to the free-quark decay, contrary to the case of the $B_d$, and $B_s$ mesons, where the effect of four-quark operators is much smaller~\cite{Lenz:2022rbq}.} In both cases we observe a rather mild dependence on $\mu_b$, even at LO-QCD.
However, the perturbative series does not seem to converge,
as the relative shift at NNLO with respect to the NLO value is larger than the one at NLO with respect to the LO result.
Furthermore, there is
a rather strong dependence on the loop order used to compute the pole masses
from the 
$\overline{\rm MS}$ mass values. In fact,
the difference between the NNLO-QCD curves in the two panels of
Fig.~\ref{fig:Gamma3-NNLO-vs-NLO-vs-LO_Pole} is much larger than the spread
due to the $\mu_b$ variation in each plot, showing that the dependence on $\mu_b$ alone is not able to capture the overall uncertainty due to the ambiguity on the value of the pole masses.
For these reasons, we consider the pole scheme as inadequate for the description of the $B_q$-meson decay rate and will not use it further in the following.

\subsubsection{\label{subsub::MSbar} \boldmath $\overline{\rm MS}$ scheme}
\begin{figure}[t]
\centering
\includegraphics[scale=0.5]{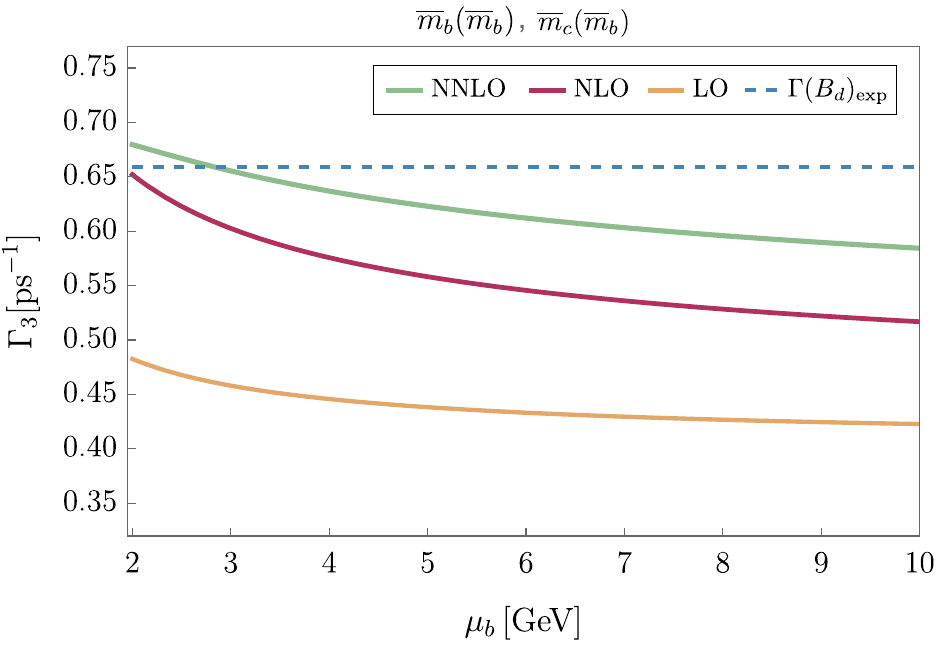} \,
\includegraphics[scale=0.5]{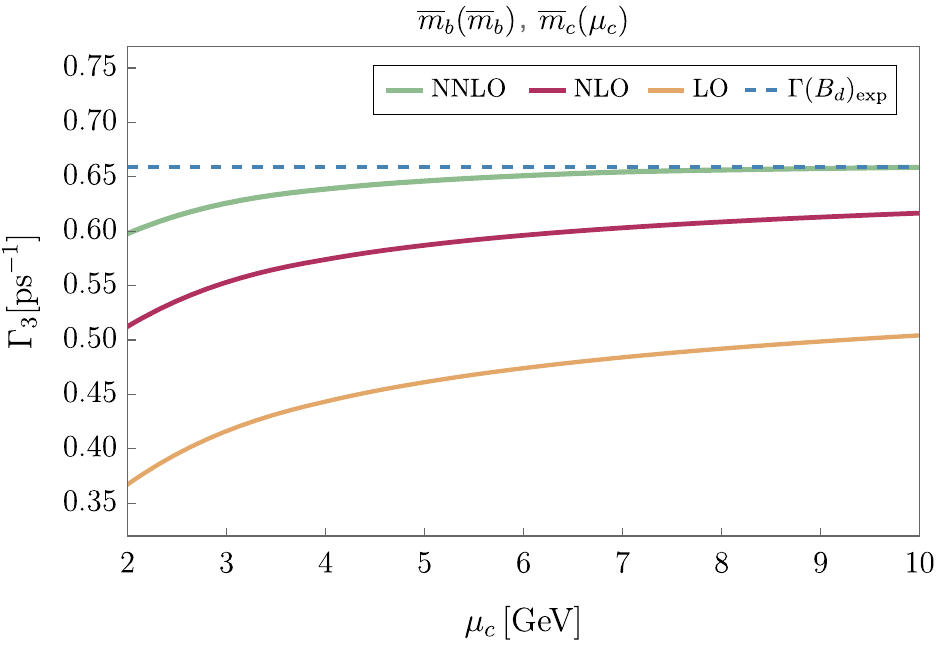}
\caption{Results at NNLO- (solid green), NLO- (solid magenta), and LO-QCD (solid orange) for $\Gamma_3$, obtained expressing both the bottom- and charm-quark masses in the $\overline{\rm MS}$ scheme, as function of the renormalisation scales $\mu_b$ (left) and $\mu_c$ (right).  The experimental value of $\Gamma(B_d)$ (dashed blue) is also shown for an approximate reference.}
\label{fig:Gamma3-NNLO-vs-NLO-vs-LO_MSbar}
\end{figure}

Next, we consider the case in which both the bottom and the charm quark masses are renormalised in the $\overline{\rm MS}$  
scheme. Setting $\mu_c = \mu_b = \overline{m}_b (\overline{m}_b)$, and using the input values in Eqs.~\eqref{eq:mb_MSbar} and~\eqref{eq:mc-MS-mbMS-SL-fit}, we obtain the following result for $\Gamma_3$, namely
\begin{eqnarray}
\Gamma_3 & = & \Gamma_0 \biggl[5.497 +  \bigl(1.336 + \underbrace{44.46}_{\delta m_b} - \underbrace{23.53}_{\delta m_c} \bigr) \frac{\alpha_s}{\pi}
\nonumber \\
& + & 
\bigl(27.45 + \underbrace{474.6}_{\delta m_b} - \underbrace{220.5}_{\delta m_c} - \underbrace{128.9}_{\delta m_b \, \delta m_c} \bigr) \left(\frac{\alpha_s}{\pi}\right)^2
\biggr],     
\label{eq:Gamma3-decomp_MSbar}
\end{eqnarray}
where $\Gamma_0 = 0.08146 \psinv$,
and the labels $\delta m_b$ and $\delta m_c$ indicate, respectively, the contributions stemming from the conversion of the bottom and charm quark masses from the pole to the $\overline{\rm MS}$ scheme. 
The dependence of the partonic $b$-quark decay in the $\overline{\rm MS}$ scheme, on  $\mu_b$ and $\mu_c$, is shown, respectively, on the left and right panels of Fig.~\ref{fig:Gamma3-NNLO-vs-NLO-vs-LO_MSbar}, together with the experimental value of $\Gamma(B_d)$, which, as stated before, we use as a reference for the experimental value of $\Gamma_3$. In this case, we observe large QCD corrections both at NLO and NNLO, in particular one obtains large positive contributions from the transformation of the bottom-quark mass to the $\overline{\rm MS}$ scheme which are only
partly compensated form the charm-quark contributions, and as a result the
NLO and NNLO corrections are larger than in the pole scheme.
Furthermore, in each panel,
the uncertainty bands due to the variation of $\mu_c$ and $\mu_b$
barely overlap. It should also be emphasised that for most of the parameters of the HQE there is no determination available in this scheme.
Thus, for the numerical prediction of the
total decay rates we will also not consider this renormalisation scheme.

\subsubsection{\boldmath Kinetic scheme for bottom and $\overline{\rm MS}$ scheme for charm}
\label{sub:kin+MS}
Finally, we turn to our default scenario, that is the use of the kinetic scheme for the bottom quark mass and of the $\overline{\rm MS}$ scheme for the charm quark. Using the input values given in Eqs.~\eqref{eq:mb-kin-SL-fit} and~\eqref{eq:mc-MS-mbMS-SL-fit}, 
we obtain, for $\mu_c = \mu_b = m_b^{\rm kin}$ and $\mu^{\rm cut} = 1 \GeV$, the following decomposition of $\Gamma_3$ up to NNLO:
\begin{eqnarray}
\Gamma_3 & = & \Gamma_0 \biggl[5.997 + \bigl(1.864 + \underbrace{14.88}_{\delta m_b} - \underbrace{22.17}_{\delta m_c} \bigr) \frac{\alpha_s}{\pi} 
\nonumber \\
& + & 
\bigl(30.67 + \underbrace{216.8}_{\delta m_b} - \underbrace{220.8}_{\delta m_c} - \underbrace{36.59}_{\delta m_b \, \delta m_c} \bigr) \left(\frac{\alpha_s}{\pi}\right)^{\!\!2\,}
\biggr],     
\label{eq:Gamma3-decomp}
\end{eqnarray}
where $\Gamma_0 = 0.1223 \psinv$. The contributions stemming from the conversion of the bottom- and charm-quark masses from the pole to the kinetic and $\overline{\rm MS}$ schemes, respectively, are labeled by $\delta m_b$ and $\delta m_c$. Note that Eq.~\eqref{eq:Gamma3-decomp} also includes the perturbative contributions $[\mu_\pi^2(1 {\rm GeV})]_{\rm pert}$ and $[\rho_D^3(1 {\rm GeV})]_{\rm pert}$ given in Eqs.~\eqref{eq:mupi-kin} and~\eqref{eq:rhoD-kin}, as it is commonly done in the literature, see e.g.\ Ref.~\cite{Finauri:2023kte}.
\begin{figure}[t]
\centering
\includegraphics[scale=0.5]{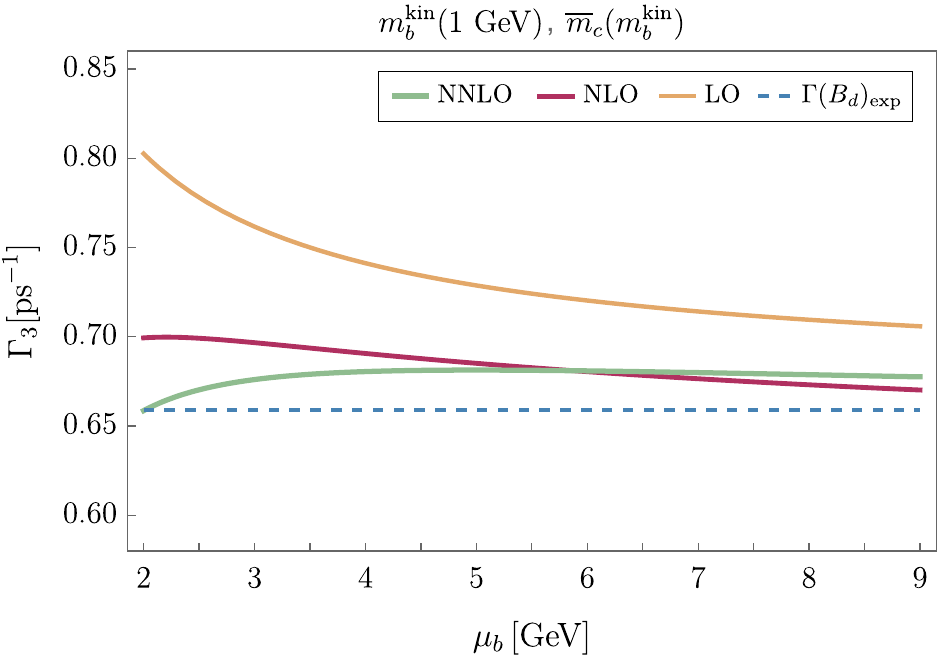} \,
\includegraphics[scale=0.5]{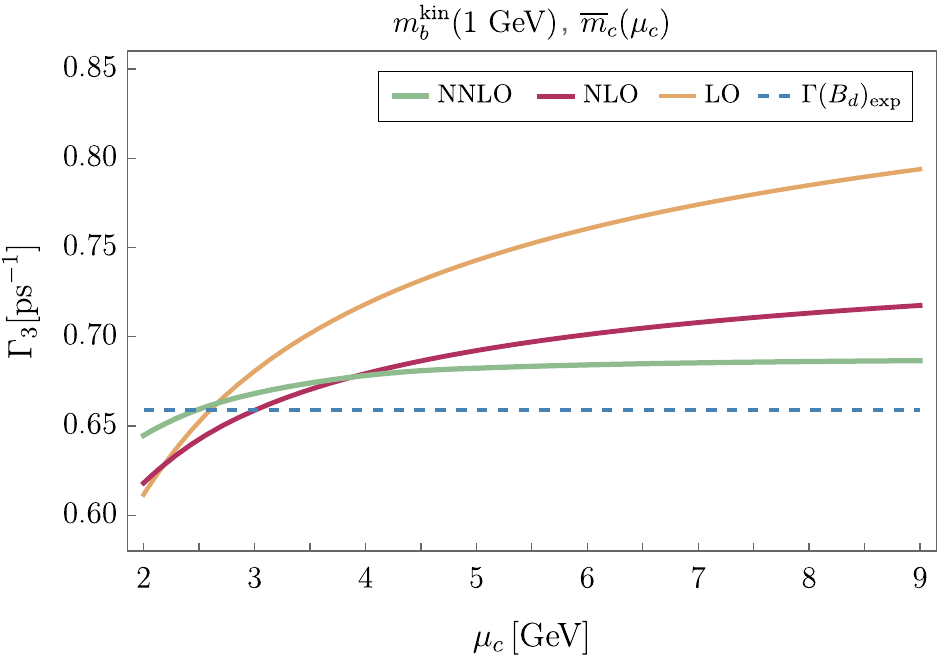} \\
\includegraphics[scale=0.5]{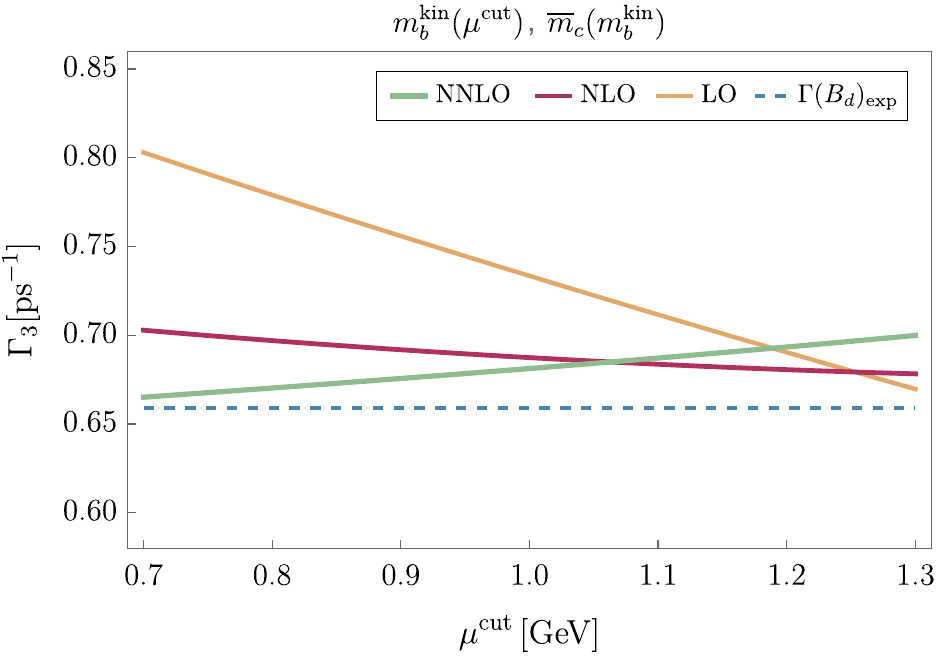} \,
\includegraphics[scale=0.5]{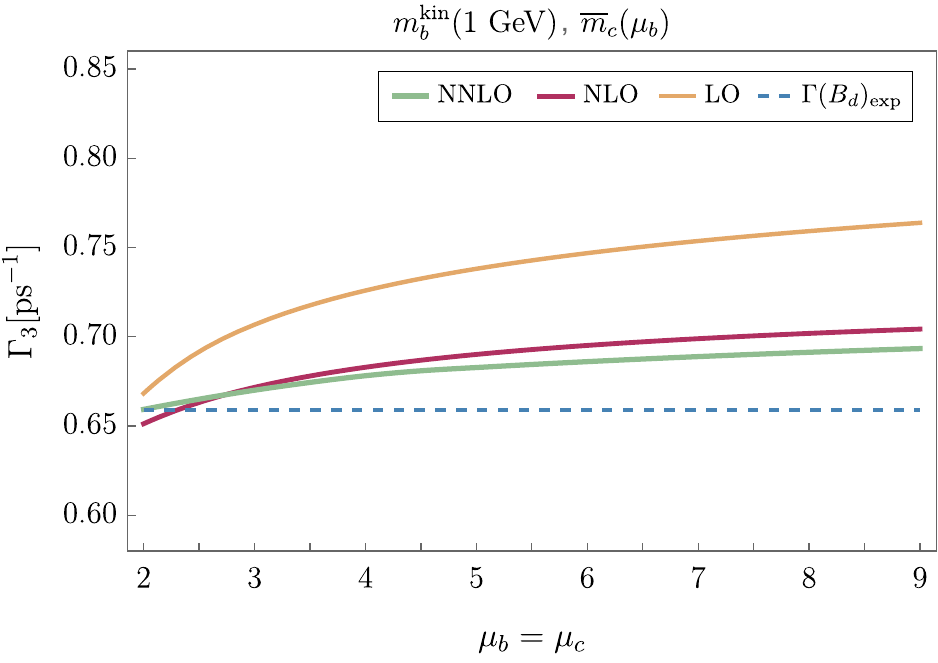} 
\caption{Results at NNLO- (solid green), NLO- (solid magenta), and LO-QCD (solid orange) for $\Gamma_3$, obtained expressing the $b$-quark mass in the kinetic scheme and the $c$-quark in the $\overline{\rm MS}$, as function of the renormalisation scales $\mu_b$ (top left), $\mu_c$ (top right), and $\mu^{\rm cut}$ (bottom left). The particular case $\mu_c = \mu_b$ is shown for completeness (bottom right). In each plot, the experimental value of $\Gamma(B_d)$ (dashed blue) is also shown for an approximate reference. }
\label{fig:Gamma3-NNLO-vs-NLO-vs-LO_kin-MSbar}
\end{figure}

The dependence of $\Gamma_3$ in Eq.~\eqref{eq:Gamma3-decomp}, on the renormalisation scales $\mu_b, \mu_c$ and  $\mu^{\rm cut}$ is depicted, respectively, on the top left, top right and bottom left panels of Fig.~\ref{fig:Gamma3-NNLO-vs-NLO-vs-LO_kin-MSbar}, together with the experimental result of~$\Gamma(B_d)$ which we use as an approximate reference for the experimental value of $\Gamma_3$. Note that on the bottom right plot of Fig.~\ref{fig:Gamma3-NNLO-vs-NLO-vs-LO_kin-MSbar} we also show the dependence of $\Gamma_3$ on the renormalisation scale $\mu_b$ in the particular case of $\mu_c = \mu_b$.
From the above plots we see how in this scheme the perturbative expansion appears well behaving, with further QCD corrections being smaller and contained within the previous orders error bands. Moreover, at NNLO-QCD both the $\mu_b$- and $\mu_c$-scale dependence is significantly reduced. Note that from the bottom left plot of Fig.~\ref{fig:Gamma3-NNLO-vs-NLO-vs-LO_kin-MSbar}, this might seem not to be the case for the dependence on $\mu^{\rm cut}$, which appears stronger at NNLO- that at NLO-QCD. This is, however, an artifact of having included in the expression for $\Gamma_3$ also the perturbative contributions $[\mu_\pi^2(\mu^{\rm cut})]_{\rm pert}$ and $[\rho_D^3(\mu^{\rm cut})]_{\rm pert}$ in Eqs.~\eqref{eq:mupi-kin} and~\eqref{eq:rhoD-kin}. In fact, once the remaining corrections due to $\mu_\pi^2$ and $\rho_D^3$ in the kinetic scheme are also included, the $\mu^{\rm cut}$ dependence of the total decay width becomes significantly more stable at NNLO than at NLO, cf.\ e.g.\ the bottom left plot of Fig.~\ref{fig:Bu-kin-MS}. For the above reasons, and also taking into account that the values of many parameters of the HQE i.e.\ $V_{cb}$, $m_b$, $\mu_\pi^2$, $\mu_G^2$, and $\rho_D^3$ have been extracted using the kinetic scheme for the bottom-quark and the $\overline{\rm MS}$ scheme for the charm~\cite{Finauri:2023kte}, we consider this to be the most appropriate scenario to carry out our final analysis.  
 
\subsection{Total decay rates at NNLO-QCD}
\label{subsec:total-widths}

In this section we present our predictions for the total decay widths of the
$B^+$, $B_d$, and $B_s$ mesons, obtained 
using the kinetic scheme for the bottom-quark mass and the $\overline {\rm MS}$ scheme for the charm quark. As for the corrections included in our results, we stress that, for consistency, we use everywhere the same accuracy for the semileptonic modes as the one currently available for the non-leptonic channels. Hence, at leading power, the known N$^3$LO-QCD corrections to the semileptonic $b\to c \ell^- \bar \nu_\ell$ decay~\cite{Fael:2020tow} are not included in our central values, as well as the corresponding NLO-QCD corrections at order $1/m_b^2$~\cite{Alberti:2013kxa, Mannel:2014xza, Mannel:2015jka} and $1/m_b^3$~\cite{Mannel:2019qel, Mannel:2021zzr, Moreno:2022goo}. The QED corrections to the $b\to c \ell^- \bar \nu_\ell$ mode, recently completely determined in Ref.~\cite{Bigi:2023cbv} are also neglected.
This argument applies also to the partial NLO-QCD contributions to the chromo-magnetic operator due to non-leptonic decays, which are known only for the $b\to c \bar u d$ mode~\cite{Mannel:2024uar}. However, the effect of including all the available corrections on our predictions is discussed later in the text.

\begin{figure}[t]
    \centering
    \includegraphics[scale=0.51]{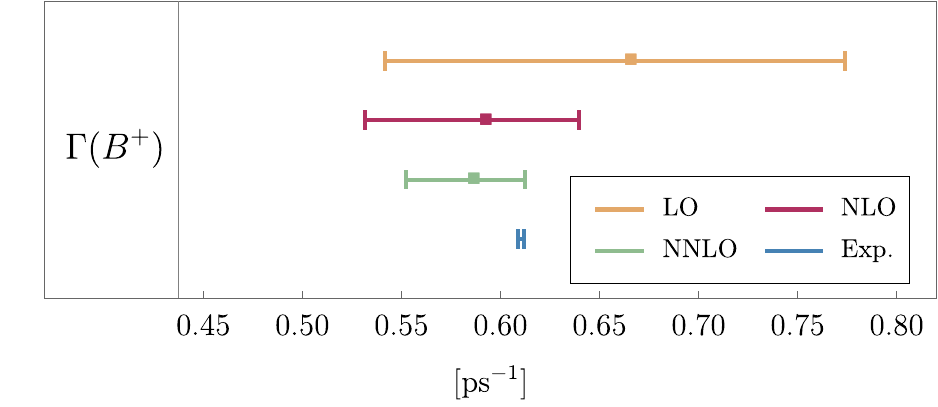}
    \,
    \includegraphics[scale=0.51]{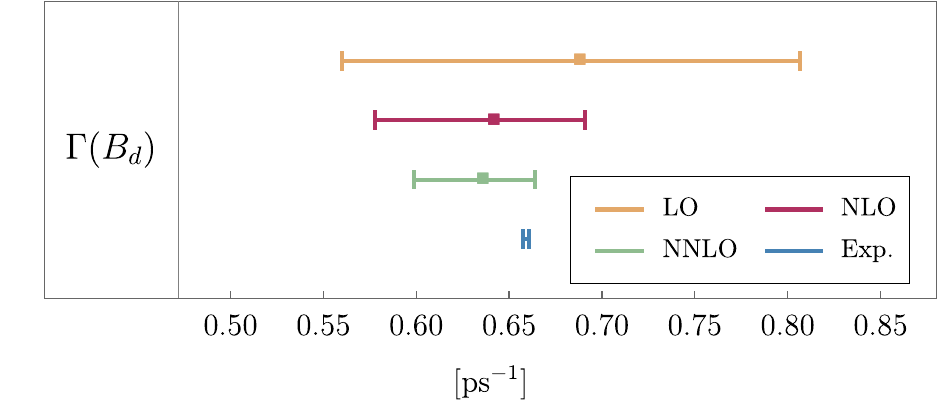}
    \\[1mm]
    \includegraphics[scale=0.51]{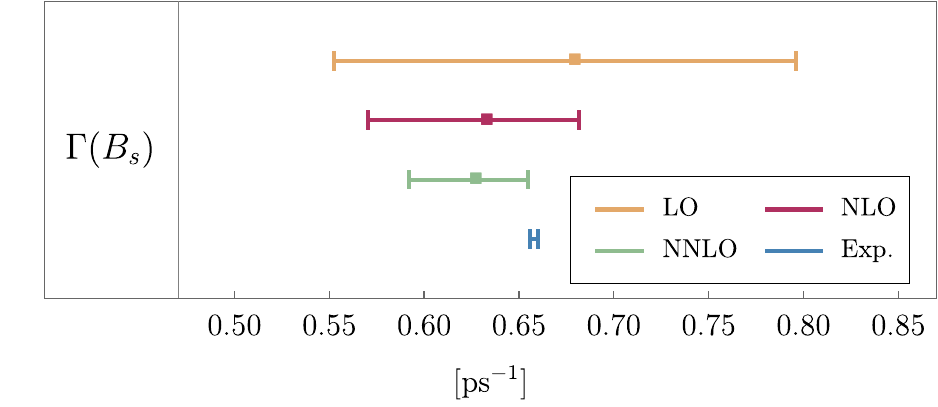}
    \caption{Comparison of the HQE predictions for the total decay widths of the $B^+$ (top left), $B_d$ (top right) and $B_s$ (bottom) mesons. In each panel, the LO- (solid orange), NLO- (solid magenta), and NNLO-QCD (solid green) results are shown together with the corresponding experimental value (solid blue).}
    \label{fig:Gamma-Bq-summary}
\end{figure}

Our predictions for the total decay widths at LO- NLO- and NNLO-QCD are summarised in Fig.~\ref{fig:Gamma-Bq-summary} and in Table~\ref{tab:Gamma-Bq-summary}, together with the corresponding experimental values. 
A comparison of the theoretical predictions for the widths of the $B^+$, $B_d$, and $B_s$ mesons as function of the renormalisation scales $\mu_b, \mu_c, \mu^{\rm cut}$ and $\mu_0$ is also shown in Fig.~\ref{fig:Bu-kin-MS}, Fig.~\ref{fig:Bd-kin-MS}, and Fig.~\ref{fig:Bs-kin-MS} of Appendix~\ref{app::plots}. The results show that, going from LO to NNLO, there is a strong reduction of the theoretical uncertainties, mainly due to the reduction of the renormalisation scale dependence when including higher order QCD corrections. This effect is particularly significant for the dependence on $\mu_c$ and $\mu^{\rm cut}$, and to a lesser extent is also present for the one on $\mu_b$. On the other hand, the $\mu_0$ dependence is mild already at LO-QCD, since it originates only from the contribution of dimension-six four-quark operators which are power suppressed. 
Note that the error bands indicated in Fig.~\ref{fig:Gamma-Bq-summary} and in Table~\ref{tab:Gamma-Bq-summary}, apart from including the parametric uncertainties and those due to the variation of the renormalisation scales, all added in quadrature, as discussed in Section~\ref{subsec:input}, also 
include an additional uncertainty to account for missing higher-order corrections. Specifically, 
we add: (a) $20\%$ of the Darwin-operator contribution to account for missing power-suppressed corrections at dimension-seven due to two-quark operators; 
(b) $20\%$ of the dimension-seven four-quark operators contribution to account for missing dimension-eight corrections; 
(c) $2\%$ of the leading dimension-three  contribution at LO-QCD to account for missing QED effects; (d) $100\%$ of the LO-QCD coefficient of the chromo-magnetic operator to account for the cancellations that arise in the corresponding Wilson coefficients at this order and which could be lifted at NLO, as it was found in the case of the $b \to c \bar u d$-channel~\cite{Mannel:2024uar}.
The additional uncertainties (a)-(d) are then again combined in quadrature.
Finally, the uncertainties due to missing higher order perturbative QCD corrections, e.g.\ due to missing N$^3$LO dimension-three corrections, are expected to be covered by the variation of the renormalisation scales. Overall, the theoretical uncertainties remain dominated by the renormalisation scales variation -- although the latter is dramatically reduced going from from LO to NNLO -- and by the values of $V_{cb}$ and $m_b^{\rm kin}$ that enter the prefactor $\Gamma_0$. It should be mentioned, however, that the current accuracy on $V_{cb}$, extracted from the fit to data on inclusive semileptonic $B$ decays ~\cite{Finauri:2023kte}, reaches a remarkable level of $\sim 1.1\%$, cf.~Eq.~\eqref{eq:Vcb-value}, thus leading only to a moderate uncertainty of $\sim2\%$ to the total decay widths.

From Fig.~\ref{fig:Gamma-Bq-summary} and Table~\ref{tab:Gamma-Bq-summary} we see that within uncertainties, there is a good agreement between our NNLO predictions for the total decay widths with the corresponding HFLAV values. 
On the other hand, it is also evident from Fig.~\ref{fig:Gamma-Bq-summary} and Table~\ref{tab:Gamma-Bq-summary}, that there is a systematic, almost universal, negative deficit of the order of $\sim - 4 \%$ at the level of the central values for the NNLO predictions with respect to the experimental data. This shift could be, in principle, accommodated by missing higher-order contributions, including e.g.\ the complete $\alpha_s$-corrections to the dimension-five chromo-magnetic operator and also QED effects.
As commented before, our results do not include all available contributions to semileptonic $b$-quark decays, such as the N$^3$LO~\cite{Fael:2020tow} and complete QED corrections~\cite{Bigi:2023cbv} at leading power. Moreover, the partial QCD contributions to the chromo-magnetic operator for non-leptonic $b$-quark decays, namely due to $b \to c \bar u d$~\cite{Mannel:2024uar} have, for consistency, also been neglected. It is, therefore, instructive to study what is the effect that the inclusion of these partial contributions would have on our results, especially in light of the currently observed small deficit with respect to data for the total widths. Specifically, if we distinguish the effects
\begin{itemize}
\item[(i)] Known QCD corrections to the semileptonic $b \to c \ell^- \bar \nu_\ell$ decay, with $\ell = e, \mu$,
\item[(ii)] QED contributions to the semileptonic $b \to c \ell^- \bar \nu_\ell$ decay, with $\ell = e, \mu$,
\item[(iii)] $\alpha_s$-corrections to $\mu_G^2$ due to the $b\to c \bar u d$ channel,
\end{itemize}
we obtain, as compared to the central values quoted in Fig.~\ref{fig:Gamma-Bq-summary} and Table~\ref{tab:Gamma-Bq-summary}, respectively, a negative shift of $\sim - 0.5 \%$ due to (i), a positive shift of $\sim + 0.7 \%$ due to (ii), and a positive shift of $\sim + 0.5 \%$ from (iii). Consequently, the net effect of including these corrections would be a positive shift of $\sim + 0.7\%$ in the central values of the total widths. This may not seem particularly large, but it should be noted that it results from including corrections from only a limited number of modes.  
Although the size of missing contributions cannot be reliably quantified in the absence of a proper calculation, we can estimate it, assuming that these corrections have the same behaviour as those currently known.
For instance, the complete QED effects in the semileptonic decay widths $b \to c \ell^- \bar \nu_\ell$ were found to be as large as $+2.31\%$ with respect to the LO-QCD partonic level contribution~\cite{Bigi:2023cbv}. It is not excluded then, that future determinations of the corresponding corrections to the tauonic channel $b \to c \tau^- \bar \nu_\tau$ and to the non-leptonic modes~\footnote{First determinations of the electro-weak (EW) corrections to the Wilson coefficients of the non-leptonic operators in the effective Hamiltonian were performed in Refs.~\cite{Gambino:2000fz, Gambino:2001au, Brod:2008ss}; as for the semileptonic modes, these corrections were also found to be positive~\cite{Bigi:2023cbv}.} could lead to a large positive shift of $1 - 2\%$ with respect to the current NNLO-QCD results for the total widths shown in Table~\ref{tab:Gamma-Bq-summary}. 
Similarly, the inclusion of the the full NLO corrections to the dimension-five contribution, once the missing corrections in the $b \to c \bar c s$ case will be computed, may ultimately yield a visible positive shift in the total decay widths of up to $+1\%$.

\begin{figure}[t]
    \centering
     \includegraphics[scale=0.51]{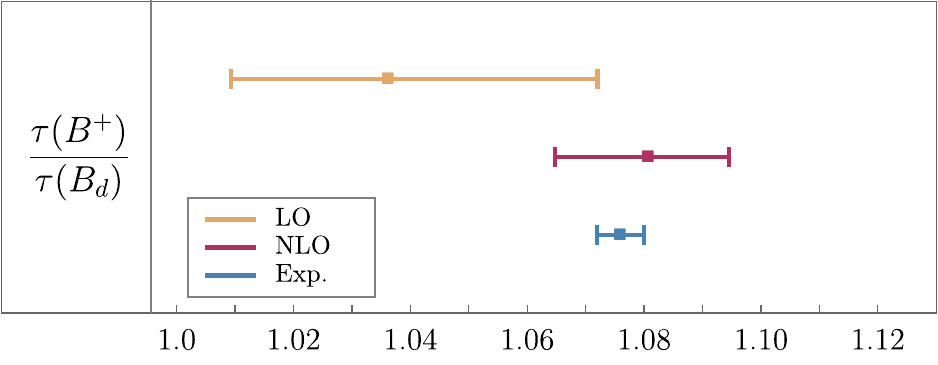}
    \,
    \includegraphics[scale=0.51]{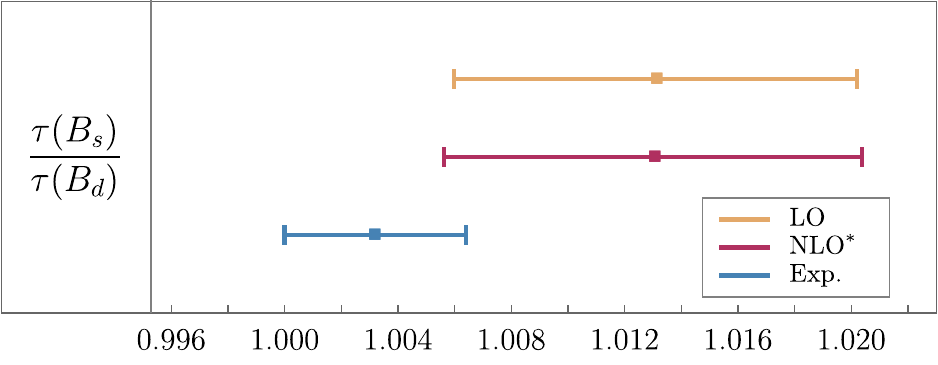}
    \caption{Comparison of the HQE predictions for the lifetime ratios $\tau (B^+)/\tau (B_d)$ (left) and $\tau (B_s)/\tau (B_d)$ (right). In each panel, the LO- (solid orange) and NLO- (solid magenta) results are shown together with the corresponding experimental value (solid blue). $^*$ Note that for $\tau (B_s)/\tau (B_d)$, the NLO-QCD result quoted is only partial as the complete $\alpha_s$-corrections to the two-quark operators contribution, which are expected to give the dominant effect at NLO, are not yet known.}
    \label{fig:Lifetime-ratios-summary}
\end{figure}

Finally, in Fig.~\ref{fig:Lifetime-ratios-summary} and Table~\ref{tab:lifetime-ratios-summary} we show our updated values for the lifetime ratios $\tau (B^+)/\tau (B_d)$ and $\tau (B_s)/\tau (B_d)$ at NLO-QCD. Note that our predictions for these ratios are obtained using the experimental data for the lifetime of the $B^+$ and $B_s$ mesons in Eq.~\eqref{eq:tau_ratio}, hence they are independent of the free $b$-quark decay and unaffected by the inclusion of the new NNLO-QCD results from Ref.~\cite{Egner:2024azu}. We stress, however, that while for $\tau (B^+)/\tau (B_d)$, the NLO value quoted reflects the current accuracy on this observable, corresponding to the NLO accuracy on the four-quark operator contributions,
for $\tau (B_s)/\tau (B_d)$, it is only partial, as indicated by the~symbol~$^*$. This follows from the fact that
the contribution of two quark operators, which is not yet completely known at NLO, exactly cancels in $\tau (B^+)/\tau (B_d)$ in the isospin limit, but not in $\tau (B_s)/\tau (B_d)$, the latter being driven by the size of the SU(3)$_F$ breaking in the corresponding non-perturbative parameters. In this regard, it is also worth emphasising that the inclusion of the $\alpha_s$-corrections to the two-quark operators contribution, once these will be fully determined, will likely have a sizable impact on the NLO prediction of $\tau (B_s)/\tau (B_d)$, as the effect of the four-quark operators is strongly suppressed in this ratio. 

From Fig.~\ref{fig:Lifetime-ratios-summary} and Table~\ref{tab:lifetime-ratios-summary}, we see that for $\tau (B^+)/\tau (B_d)$, the $\alpha_s$-corrections cause a positive shift of $\sim + 4\%$ of the LO central value, as well as a significant reduction of the uncertainties. The final result is in very good agreement with the data as is it was already found in the previous study~\cite{Lenz:2022rbq}; however note how now, with the inclusion of the updated values for the dimension-six Bag parameters~\cite{Bag-parameters-new}, the agreement with the data is even improved, cf.\ Eq.~\eqref{eq:tauBs/tauBd_HQE_old}. For $\tau (B_s)/\tau (B_d)$, on the other hand, the inclusion of the $\alpha_s$-corrections due to four-quark operators leads only to a very minor effect, so that the corresponding QCD corrections to the two-quark operators might be significant, as already commented above. We stress however that now the value quoted in Table~\ref{tab:lifetime-ratios-summary} has changed compared to the result of the previous analysis~\cite{Lenz:2022rbq} and lies in between the two scenarios considered in the latter reference. This mainly follows from having now included, differently from what was done in Ref.~\cite{Lenz:2022rbq}, the perturbative contributions to $\rho_D^3$ in the determination of the corresponding value for $\rho_D^3(B_s)$, see Eq.~\eqref{eq:rhoD-kin-Bs}, which leads to a sizable reduction of the SU(3)$_F$-breaking effects in this parameter. The small negative shift of $\sim - 5\%$ in the value of $\rho_D^3(B)$ obtained from the recent semileptonic fit \cite{Finauri:2023kte}, as compared to the previous determination~\cite{Bordone:2021oof}, also contributes to this difference\footnote{Note that, following the computation of the complete NNLO-QCD corrections to the $q^2$-spectrum~\cite{Fael:2024gyw}, the semileptonic fit has been recently updated and the new results have been presented at the workshop ``Challenges in semileptonic $B$ decay''~\cite{Finauri@challenges}.
The updated fit gives slightly smaller values for $V_{cb}$ and $\rho_D^3$, causing a minor shift of $\approx - 0.3~\%$ in our current NNLO predictions shown in Table~\ref{tab:Gamma-Bq-summary}, and of $\lesssim -0.1 \%$ in both the lifetime ratios $\tau(B^+)/\tau(B_d)$ and $\tau(B_s)/\tau(B_d)$ shown in Table~\ref{tab:lifetime-ratios-summary}.}. For completeness, we also discuss the effect of determining the lifetime ratios entirely within the HQE, that is using our predictions at NNLO-QCD for $\tau(B^+)$ and $\tau(B_s)$ in Eq.~\eqref{eq:tau_ratio}, rather than the corresponding experimental data. In this case our central values would read respectively $\tau(B^+)/\tau(B_d) = 1. 084$ and $\tau(B_s)/\tau(B_d) = 1.0137$, corresponding to a sub-percent shift with respect to the values shown in Table~\ref{tab:lifetime-ratios-summary}. 

Finally, a comment on the value of $V_{cb}$ is in order. We stress again that in our predictions we use the value of $V_{cb}$ extracted from the inclusive fit, see Eq.~\eqref{eq:Vcb-value}. The small deficit of our central values for the total decay widths compared to the experimental data, in principle, could be accommodated by a slight enhancement of $V_{cb}$. 
By performing a ``naive'' fit, i.e.\ assuming that the HQE prediction perfectly reproduces the experimental value for the total decay rates, one would obtain the following estimates 
\begin{eqnarray}
|V_{cb}| & = & 42.8^{+0.9}_{-1.3} \times 10^{-3}\,, \qquad B^+\,,\\
|V_{cb}| & = & 42.7^{+0.9}_{-1.2} \times 10^{-3}\,, \qquad B_d\,, \\ 
|V_{cb}| & = & 43.0^{+0.9}_{-1.2} \times 10^{-3}\,, \qquad B_s\,, 
\end{eqnarray}
which, as expected, are a bit larger than the value of $V_{cb}$ obtained from the inclusive fit, though consistent with the latter within uncertainties.
It should of course be kept in mind that these values are purely illustrative and a definite statement currently cannot be made since,
as already discussed above, missing perturbative and power corrections can potentially change this picture.
On the other side, using the exclusive value of $V_{cb} = (39.8 \pm 0.6) \times 10^{-3}$ \cite{PDG:2024} would lead to a stronger tension of about $\sim 3.3\sigma$ with the experimental data.


\begin{table}[t]\centering
\renewcommand{\arraystretch}{1.7}
\begin{tabular}{|C{2.8cm}||C{2.8cm}|C{2.8cm}|C{2.8cm}||C{3.0cm}|}
\hline
Observable & LO & NLO & NNLO & Exp. value \\
\hline
\hline
$\Gu [{\rm ps}^{-1}]$ 
& $0.666^{+0.108}_{-0.124} $
& $0.593^{+0.047}_{-0.061} $
& $0.587^{+0.025}_{-0.035} $
& $0.6105 \pm 0.0015 $
\\
\hline
$\Gd [{\rm ps}^{-1}]$ 
& $0.688^{+0.118}_{-0.128} $
& $0.642^{+0.049}_{-0.064} $
& $0.636^{+0.028}_{-0.037} $
& $0.6592 \pm 0.0017 $
\\
\hline
$\Gs [{\rm ps}^{-1}]$ 
& $0.680^{+0.116}_{-0.127} $
& $0.633^{+0.048}_{-0.063} $
& $0.628^{+0.027}_{-0.035} $
& $0.6579 \pm 0.0022 $
\\
\hline
\end{tabular}
\caption{Our predictions for the $B^+$-, $B_d$-, and $B_s$-meson decay widths, based on the HQE, at LO-, NLO- and NNLO-QCD, within our default scenario, that is using the kinetic scheme for the $b$-quark and the $\overline{\rm MS}$-scheme for the $c$-quark masses. The corresponding experimental values are shown for comparison in the last column.}
\label{tab:Gamma-Bq-summary}
\end{table}

\begin{table}[th]
\centering
\renewcommand{\arraystretch}{1.7}
\begin{tabular}{|c|c|c|c|}
\hline
Observable & LO & NLO & Exp. value \\
\hline
         \hline
$\tud $ 
& $1.036^{+0.036}_{-0.027} $
& $1.081^{+0.014}_{-0.016} $
& $1.076 \pm 0.004 $
\\
\hline
$\tsd $ 
& $1.0132^{+0.0070}_{-0.0072} $
& $\left[1.0131^{+0.0073}_{-0.0074}\right]^*$
& $1.0032 \pm 0.0032 $
\\
\hline 
    \end{tabular}
    \caption{Our predictions, based on the HQE, for the lifetime ratios $\tau(B^+)/\tau(B_d)$ and $\tau(B_s)/\tau(B_d)$ at LO- and NLO-QCD, in our default scenario, that is using the kinetic scheme for the $b$-quark and the $\overline{\rm MS}$-scheme for the $c$-quark masses. The corresponding experimental values are shown for comparison in the last column. $^*$ Note that for $\tau(B_s)/\tau(B_d)$, the NLO-QCD result quoted is only partial as the complete $\alpha_s$-corrections to the two-quark operators contribution, which are expected to give the dominant effect at NLO, are not yet known.}
    \label{tab:lifetime-ratios-summary}
\end{table}

\subsection{The semileptonic branching fractions within the HQE} 
\label{subsec:sl-Brs}
In this section we present our predictions, obtained entirely within the HQE i.e.\ without using the experimental values of $\Gamma (B_q)$, for the semileptonic branching fraction ${\cal B}_{sl}$ of the three light $B_q$ mesons, defined as
\begin{eqnarray}
    {\cal B}_{\rm sl} (B_q) & = & \frac{\Gamma_{\rm sl} (B_q)}{\Gamma (B_q)}\,,
    \label{eq:Bsl}
\end{eqnarray}
where $\Gamma_{\rm sl} (B_q) \equiv \Gamma (B_q \to X_c \, e^- \bar \nu_e) = \Gamma (B_q \to X_c \, \mu^- \bar \nu_\mu)$.
To compute $\Gamma_{\rm sl}(B_q)$ we use the expression implemented in the package {\tt kolya}~\cite{Fael:2024fkt}, which includes N$^3$LO-QCD corrections to the partonic decay rate~\cite{Fael:2020tow}, as well as NLO corrections up to $1/m_b^3$~\cite{Mannel:2021zzr}. 
For the denominator of Eq.~\eqref{eq:Bsl}, we determine $\Gamma(B_q)$ as discussed in the previous section, also adding, for consistency with the numerator, all available corrections to the semileptonic modes $B_q \to X_c \ell^- \bar \nu_\ell$, with $\ell = e, \mu$.
Note that we do not re-expand the ratio in Eq.~\eqref{eq:Bsl} in $\alpha_s$, however, we have checked that the numerical difference in doing so is small. 
Our results read 
\begin{eqnarray}
{\cal B}_{\rm sl} (B^+) & = & (11.46^{+0.47}_{-0.32}) \%\,, 
\label{eq:Bsl-B+}
\\  
{\cal B}_{\rm sl} (B_d) & = & (10.57^{+0.47}_{-0.27}) \%\,, 
\label{eq:Bsl-Bd}
\\
{\cal B}_{\rm sl} (B_s) & = & (10.52^{+0.50}_{-0.29}) \%\,.
\label{eq:Bsl-Bs}
\end{eqnarray}
We emphasise that the above values do not include QED corrections. In fact, on the one hand, the effect of photon radiation at scales below $\mu \sim m_b$ is subtracted from the corresponding experimental data
using PHOTOS~\cite{Davidson:2010ew}. 
On the other hand, the EW corrections at scales above $\mu \sim m_b$ are expected to be of similar size for both the non-leptonic and semileptonic decays, and thus, to a large extent, to cancel out in the ratio, cf.~eq.~\eqref{eq:Bsl}. 

Our result for $B^+$ and $B_d$ can be compared to the experimental average 
\begin{equation}
\mathcal{B}_\mathrm{sl}(B^\pm/B_d \text{ averaged}) = (10.48 \pm 0.13) \%
\end{equation}
reported in Ref.~\cite{Bernlochner:2022ucr} and based on an extrapolation to the
full phase space of the branching ratio measurements
performed at CLEO~\cite{CLEO:2004stg}, Babar~\cite{BaBar:2009zpz,BaBar:2016rxh} and Belle~\cite{Belle:2006kgy}.
The results can also be compared with the value $\mathcal{B}_\mathrm{sl}(B^\pm/B_d \text{ averaged}) = (10.63 \pm 0.15) \%$, 
extrapolated from the global fit in Ref.~\cite{Finauri:2023kte}.

\begin{figure}[th]
    \centering
    \includegraphics[width=0.5\textwidth]{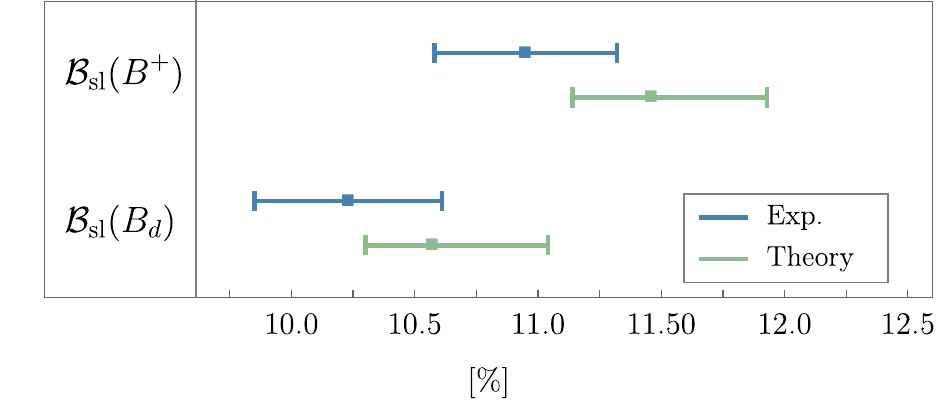}
    \caption{Our predictions for the semileptonic branching fractions of the $B^+$ and $B_d$ mesons within the HQE (solid green) compared with the measurements by Belle~\cite{Belle:2006kgy} (solid blue).}
\label{fig:Bsl_comparison}
\end{figure}

The Belle collaboration measured the independent $B^+$ and $B^0$ partial branching fractions with an electron energy greater than $E_\mathrm{cut}=0.4$~GeV in Ref.~\cite{Belle:2006kgy}.
We can extrapolate these measurements with a cut on the lepton energy to the full phase space with the correction factor $\Delta(E_\mathrm{cut})$, such that
\begin{eqnarray}
    B_{\rm sl} (B_q) = 
    \Delta(E_\mathrm{cut}) B_{\rm sl} (B_q \to X_c l^- \bar \nu_l, E_e > E_\mathrm{cut}).
\end{eqnarray}
Using the local OPE up to $1/m_b^3$, we calculate $\Delta(E_\mathrm{cut})$ using the expressions given in \texttt{kolya}~\cite{Fael:2024fkt} and obtain:
\begin{equation}
    \Delta(E_\mathrm{cut}=0.4\, \mathrm{GeV}) = 1.015 \pm 0.001,
    \label{eqn:DeltaEcut}
\end{equation}
where the uncertainty arises mainly from 
the parametric uncertainties of the HQE parameters from Ref.~\cite{Finauri:2023kte}.
The uncertainty due to the variation of the renormalization scale of $\alpha_s$ is negligible.
The available measurements of Belle, extrapolated to the full region using the correction factor in Eq.~\eqref{eqn:DeltaEcut}, are
\begin{align}
    {\cal B}_{\rm sl}^\mathrm{Belle} (B^+) &= (10.95 \pm 0.37 ) \%, \\
    {\cal B}_{\rm sl}^\mathrm{Belle} (B_d) &= (10.23 \pm 0.38 ) \%. 
\end{align}
The branching ratios in Eqs.~\eqref{eq:Bsl-B+} and~\eqref{eq:Bsl-Bd} are compared with the measurements by Belle in Fig.~\ref{fig:Bsl_comparison}. Our predictions are in agreement with the experimental determinations within errors
and are also comparable in precision.

The comparison of semileptonic branching fractions presented in this section 
serves as a test of both the HQE and QCD.\footnote{Note that historically the theory determination of the semileptonic branching fractions yielded values considerably above the experimental ones \cite{Bigi:1993fm} -- this appears settled now.} 
Since the dependence on $|V_{cb}|$ in the numerator and denominator of 
Eq.~\eqref{eq:Bsl} nearly cancels out when both are calculated within 
the framework of the HQE, the ratio $\mathcal{B}_\mathrm{sl}$ 
becomes an observable independent on $|V_{cb}|$. Moreover, also the dependence on the fifth power of the $m_b$ quark mass drops out in the ratio, so that these observables become mostly sensitive to the value of $\alpha_s$ and of the HQE parameters. Hence, the semileptonic branching ratios could, in principle, also be used as an additional independent way to determine the strong coupling $\alpha_s$. However, because of the small deficit at the level of the central values and the importance of missing higher-order corrections, as discussed in Section~\ref{subsec:total-widths}, we refrain from performing such a fit in the current work and postpone it to a future study.

\section{Discussion and conclusions}
\label{sec:conclusion}

In this work we have performed an up-to-date study of the total decay widths of the $B^+$, $B_d$, and $B_s$ mesons within the HQE, implementing for the first time the recently determined NNLO-QCD corrections to non-leptonic $b$-quark decays from Ref.~\cite{Egner:2024azu}. Furthermore, we have updated the results for the lifetimes ratios $\tau (B^+)/\tau (B_d)$ and $\tau (B_s)/\tau (B_d)$, as compared to the previous work~\cite{Lenz:2022rbq}, and presented new predictions for the semileptonic branching ratios of all three mesons, obtained 
entirely within the HQE. 
Our main results are summarized in Figs.~\ref{fig:Gamma-Bq-summary},~\ref{fig:Lifetime-ratios-summary} and~\ref{fig:Bsl_comparison}, in Tables~\ref{tab:Gamma-Bq-summary} and~\ref{tab:lifetime-ratios-summary}, and in Eqs.~\eqref{eq:Bsl-B+} to~\eqref{eq:Bsl-Bs}.
The new $\alpha_s^2$-corrections to the non-leptonic $b$-quark width have also allowed us to investigate both the perturbative convergence and the mass scheme dependence of $\Gamma_3$, the leading term in the HQE. The corresponding results obtained using the pole scheme for both the bottom and charm quark masses, the $\overline{\rm MS}$ scheme, and the kinetic scheme for the bottom and the $\overline{\rm MS}$ for charm are shown, respectively, in Figs.~\ref{fig:Gamma3-NNLO-vs-NLO-vs-LO_Pole},~\ref{fig:Gamma3-NNLO-vs-NLO-vs-LO_MSbar} and~ \ref{fig:Gamma3-NNLO-vs-NLO-vs-LO_kin-MSbar}. 

The NNLO predictions for the total decay widths are found to be in good agreement, within uncertainties, with the experimental values, although a systematic negative shift of $\sim -4\%$ at the level of the central values is found for all three observables. The theoretical uncertainties appear significantly reduced with the inclusion of the NNLO-QCD corrections and now reach the order of only few percent, revealing the importance that subleading effects like QCD corrections to power suppressed contributions and QED effects will have on improving the present theoretical precision. The latter, in fact is still dominant compared to the corresponding experimental one, which currently is at the sub-percent level.

As for the lifetime ratios, our result for $\tau(B^+)/\tau(B_d)$ 
is in very good agreement with the experimental data, although the theoretical errors, dominated by the uncertainty on the values of the four-quark non-perturbative inputs, are still sizeable compared to the current experimental precision. For $\tau(B_s)/\tau(B_d)$, the updated HQE value agrees with the data within errors, and the size of the theoretical uncertainties is also comparable to the experimental ones, albeit still a factor two larger. In this case, the main source of uncertainty comes from the value of the two-quark operator matrix elements and in particular from the size of SU(3)$_F$-breaking effects.

We find good agreement with the data also for the semileptonic branching fractions, however the theory predictions appear systematically larger as consequence of the corresponding deficit in the total decay widths. The computation of these observables entirely within the HQE allows to obtain a prediction which is independent of $V_{cb}$ and also less dependent on the value of $m_b$. In this case, in fact, the theoretical precision is higher and comparable with the current experimental one.

In light of the above results, in particular given the observed deficit in the central value of the NNLO results for the total decay widths, and the current theoretical uncertainties which amount to a few percent,
the following list of corrections would be needed in order to further improve the theoretical predictions in these observables:
\begin{itemize}
    \item[$\diamond$] Complete NNLO corrections due to the mixed contribution of the current-current and penguin operators in the effective Hamiltonian. These effects are currently known at the NLO accuracy~~\cite{Krinner:2013cja}, and the latter should be extended to the next order in $\alpha_s$ to obtain the full NNLO result for $\Gamma_3$.

    \item[$\diamond$] Complete QED corrections, including those to non-leptonic $b$-quark decays. 
    These contributions have been completely determined for semileptonic modes, in case of massless leptons, in the recent work~\cite{Bigi:2023cbv}, while a computation of the full QED effects in the $b \to c \tau^- \bar \nu_\tau$ channel as well as in the non-leptonic modes is still missing. 

    \item[$\diamond$] Complete $\alpha_s$-corrections to the dimension-five chromo-magnetic operator contribution in the HQE. These corrections are known for all semileptonic modes~\cite{Alberti:2013kxa, Mannel:2014xza, Mannel:2015jka, Moreno:2022goo}, and have been recently determined also for the $b \to c \bar u d$ decay~\cite{Mannel:2024uar}. However, they are still missing for the $b \to c \bar{c} s$ channel as well as for the remaining CKM suppressed non-leptonic modes.

    \item[$\diamond$] First calculation of $\alpha_s$-corrections to the Darwin operator contribution in the case of non-leptonic decays. These corrections are known for all semileptonic -- including tauonic  modes~\cite{Mannel:2019qel, Mannel:2021zzr, Moreno:2022goo} -- however, they are completely missing for the non-leptonic case.

    \item[$\diamond$] First computation of the dimension-seven two-quark operator contributions for non-leptonic modes. These corrections are currently known only in the semileptonic case~\cite{Dassinger:2006md,Mannel:2010wj,Mannel:2023yqf}.
\end{itemize}

The last three points are also of crucial importance for improving the theoretical determinations of the lifetime ratios $\tau (B^+)/\tau (B_d)$ and $\tau (B_s)/\tau (B_d)$. In this respect, the HQE predictions will also significantly benefit from future lattice determinations of the dimension-six Bag parameters, see e.g.\ Refs.~\cite{Lin:2022fun, Black:2024iwb} for preliminary studies in this direction, as well as from a first principle determination of the dimension-seven Bag parameters.

To conclude, we also provide a perspective to further extend the study performed in this work. In fact, given the current NNLO precision in the total decay widths of $B$ mesons, it would be interesting to perform, in future, a combined fit which includes these observables in addition to the currently used semileptonic $B$ decay rates, to further constrain the value of $V_{cb}$ and also obtain a consistence check of the theory framework. Finally, it would also be very instructive to study the effect that the inclusion of the NNLO-QCD corrections to the partonic decay rate $\Gamma_3$ would have on the HQE  predictions of $D$-mesons lifetimes, especially in light of the currently large theoretical uncertainties~\cite{King:2021xqp, Gratrex:2022xpm}.  

\section*{Acknowledgments}
AL wishes to thank Ulrich Haisch for useful correspondence and MLP thanks Gael Finauri for helpful communication regarding the analysis of Ref.~\cite{Finauri:2023kte}.
We thank Matthew Black, Martin Lang and Zachary Wüthrich for providing the new numerical results for the Bag parameters \cite{Bag-parameters-new} before publication. 
This research was supported by the Deutsche
Forschungsgemeinschaft (DFG, German Research Foundation) under grant 396021762 --- TRR 257 ``Particle Physics Phenomenology after the Higgs Discovery''. The work of MLP received funding from the DFG - project number 500314741. MLP is grateful to the Mainz Institute for Theoretical Physics~(MITP) of the
Cluster of Excellence {\it PRISMA+} (Project ID 390831469), for its hospitality and its partial support during the completion of this work.
AR acknowledges the partial support by the Cluster of Excellence {\it ORIGINS} (EXC 2094, Project ID 390783311) funded by the German Research Foundation (DFG) within the Germany Excellence Strategy. 
The work of M.F. was in part supported by the European 
Union’s Horizon 2020 research and innovation
program under the Marie Sk\l{}odowska-Curie grant agreement
No.~101065445 - PHOBIDE. 
K.S. has received support by the European Research Council (ERC) under the European Union’s Horizon 2020 research and innovation programme grant agreement 101019620 (ERC Advanced Grant TOPUP)
and the UZH Postdoc Grant, grant no.~[FK-24-115].

\newpage
\appendix

\section{\label{app::plots} Scale dependence of the total decay widths}
\begin{figure}[ht]
    \centering
    \includegraphics[scale=0.42]{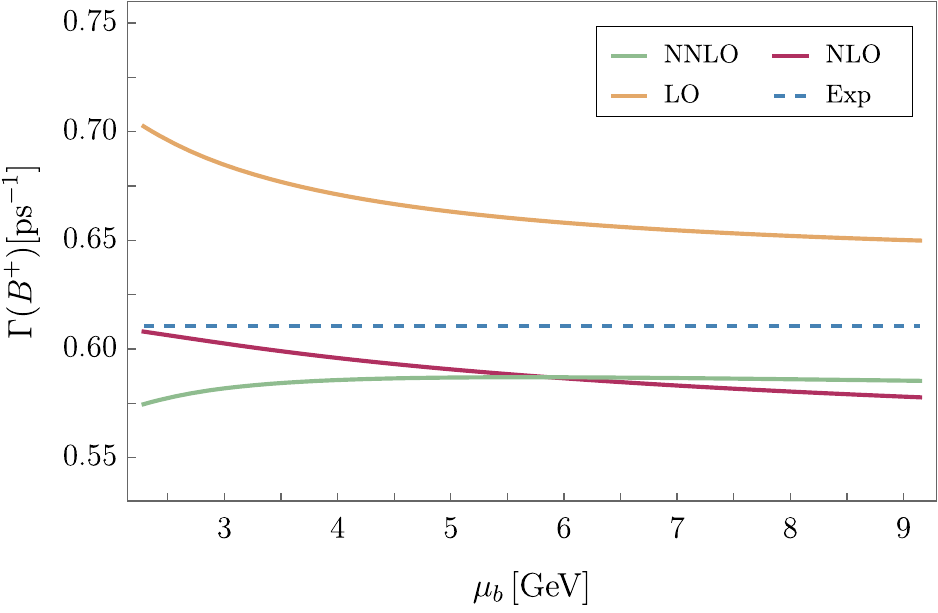} \quad
    \includegraphics[scale=0.42]{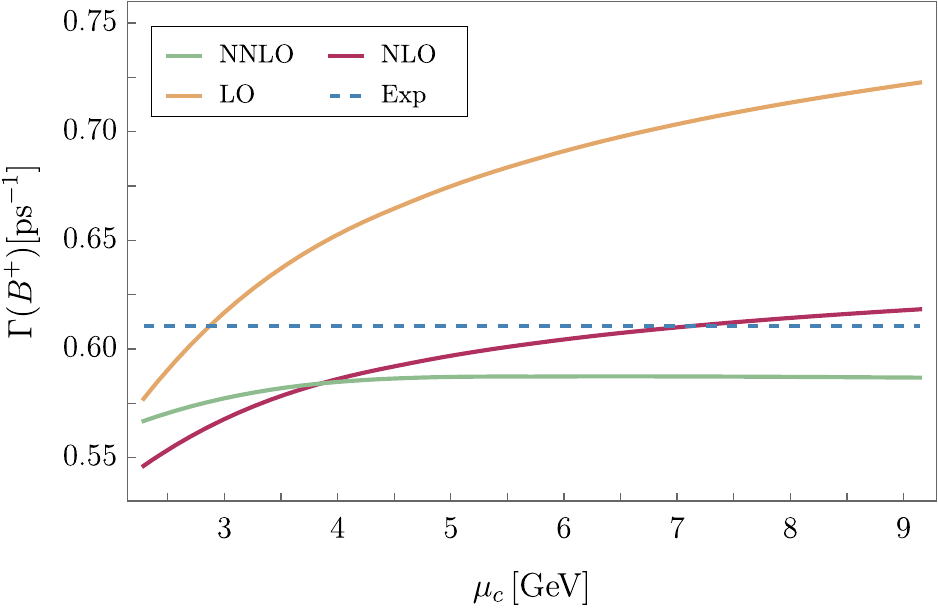}
    \includegraphics[scale=0.42]{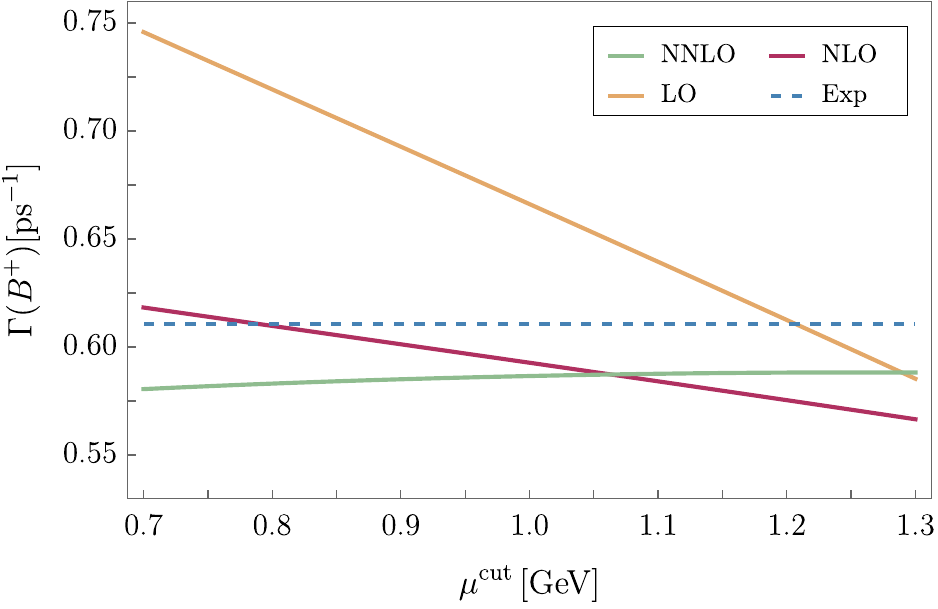} \quad
    \includegraphics[scale=0.42]{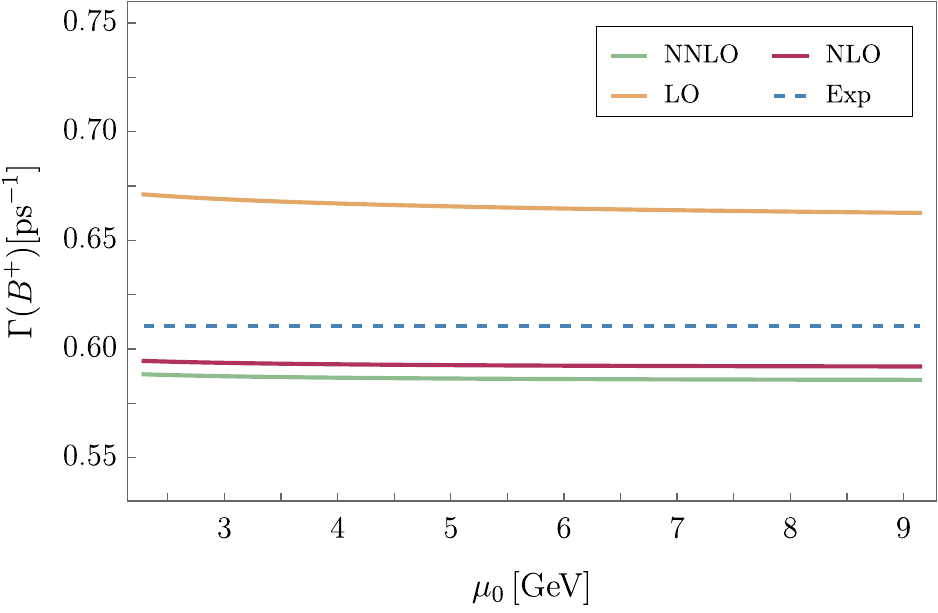}
    \caption{Dependence at NNLO- (solid green), NLO- (solid magenta), and LO-QCD (solid orange) of $\Gamma(B^+)$ on the renormalisation scales $\mu_b$ (top left), $\mu_c$ (top right), $\mu^{\rm cut}$ (bottom left), $\mu_0$ (bottom right), together with the corresponding experimental value (dashed blue).}
    \label{fig:Bu-kin-MS}
\end{figure}

\begin{figure}[ht]
    \centering
    \includegraphics[scale=0.42]{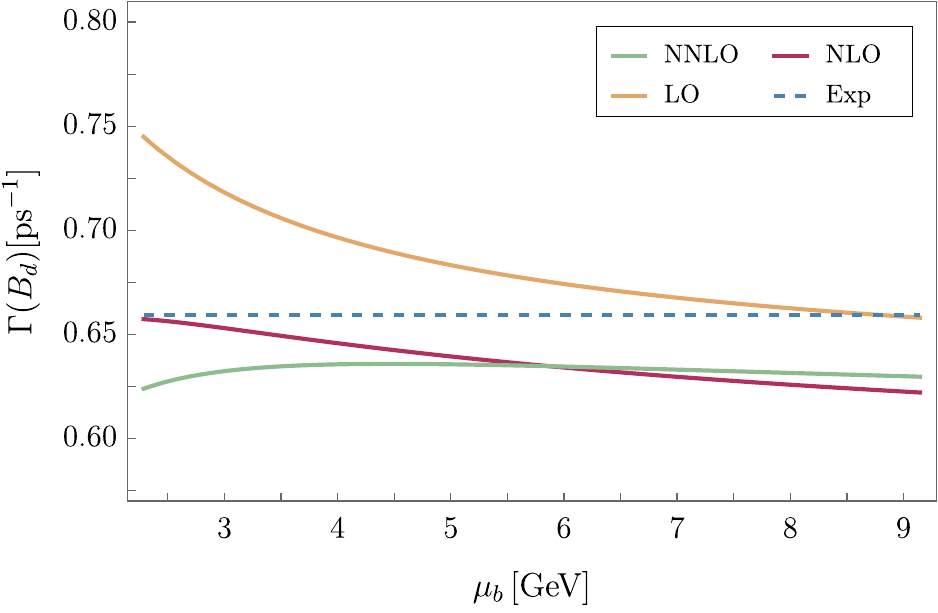} \quad
    \includegraphics[scale=0.42]{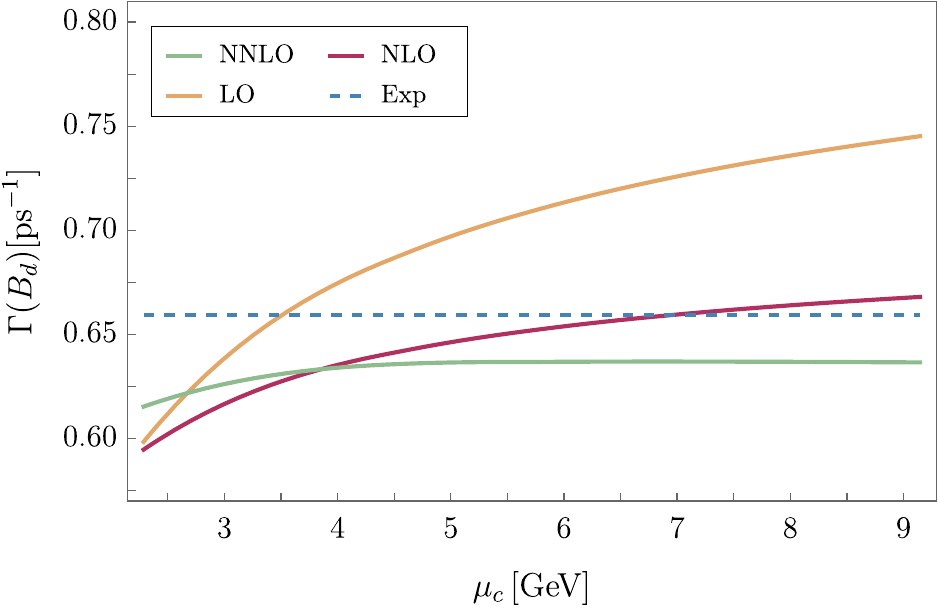}
    \includegraphics[scale=0.42]{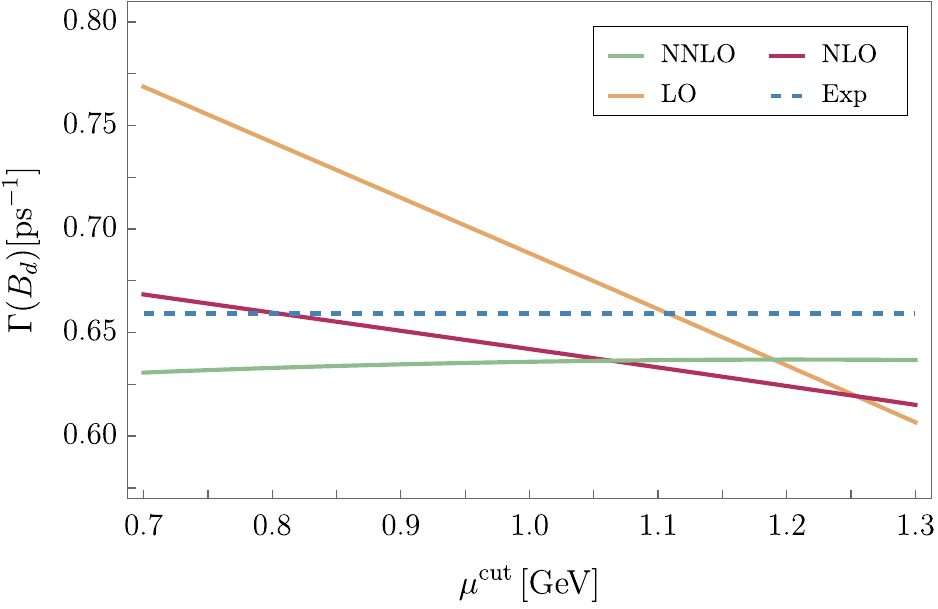} \quad
    \includegraphics[scale=0.42]{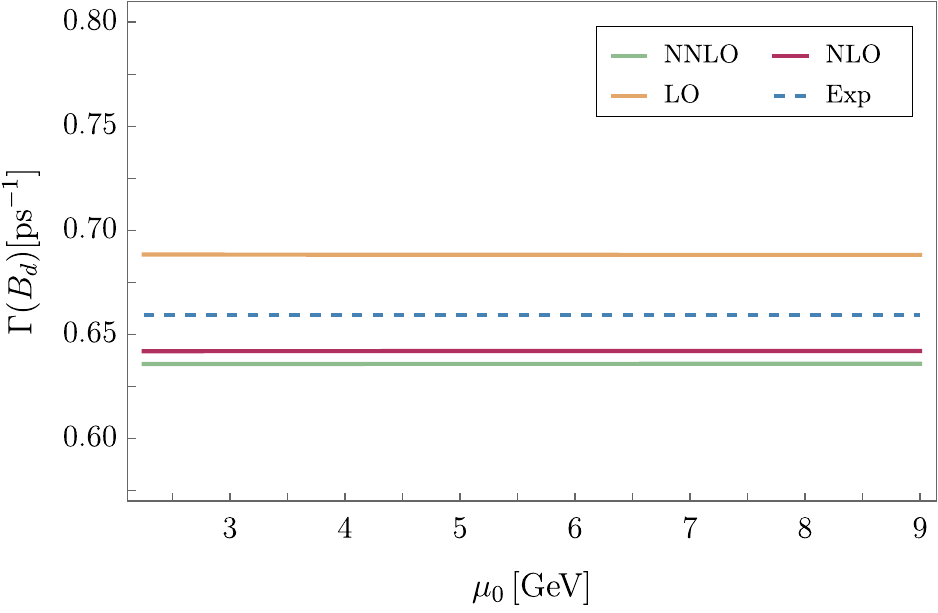}
    \caption{Dependence at NNLO- (solid green), NLO- (solid magenta), and LO-QCD (solid orange) of $\Gamma(B_d)$ on the renormalisation scales $\mu_b$ (top left), $\mu_c$ (top right), $\mu^{\rm cut}$ (bottom left), $\mu_0$ (bottom right), together with the corresponding experimental value (dashed blue).}
    \label{fig:Bd-kin-MS}
\end{figure}

\begin{figure}[ht]
    \centering
    \includegraphics[scale=0.42]{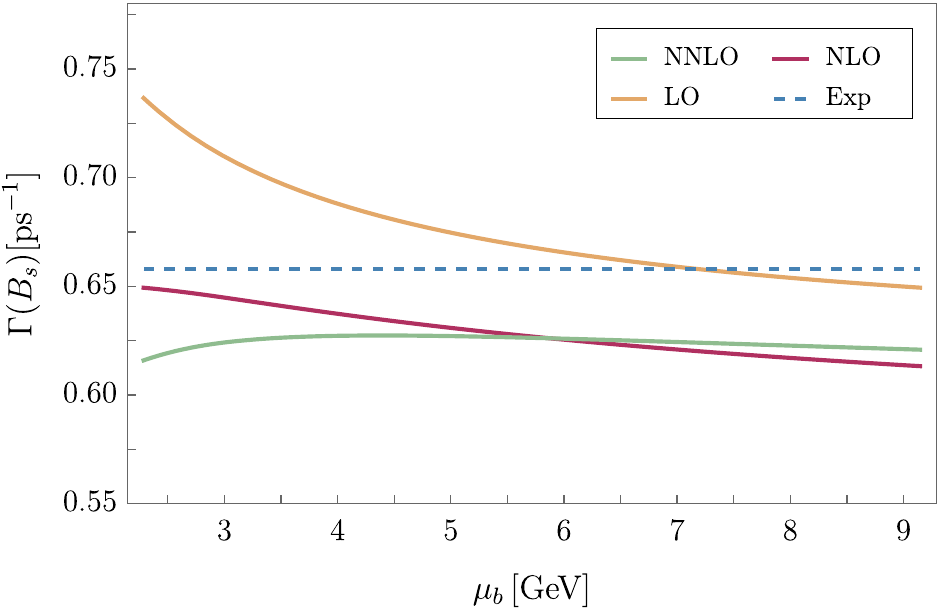} \quad
    \includegraphics[scale=0.42]{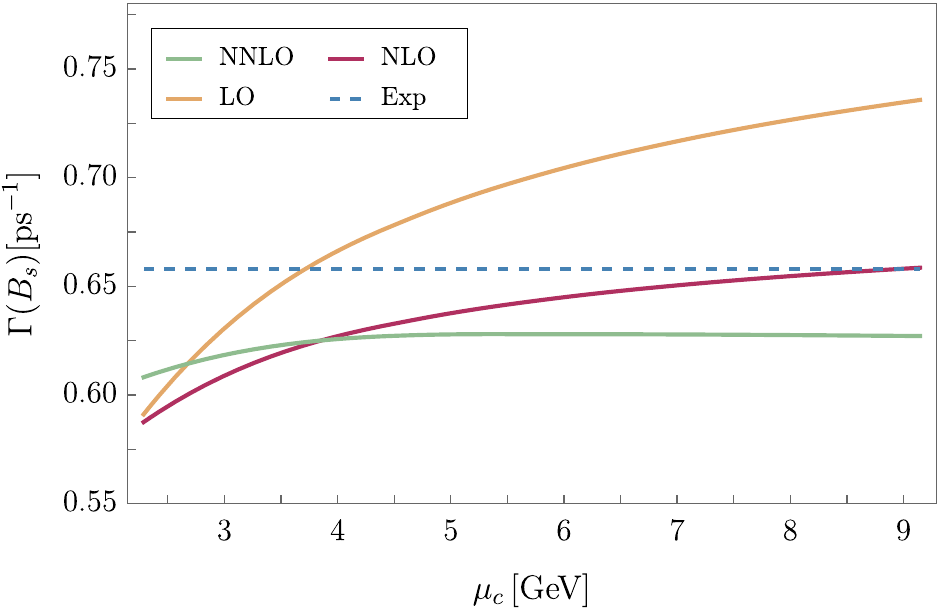}
    \includegraphics[scale=0.42]{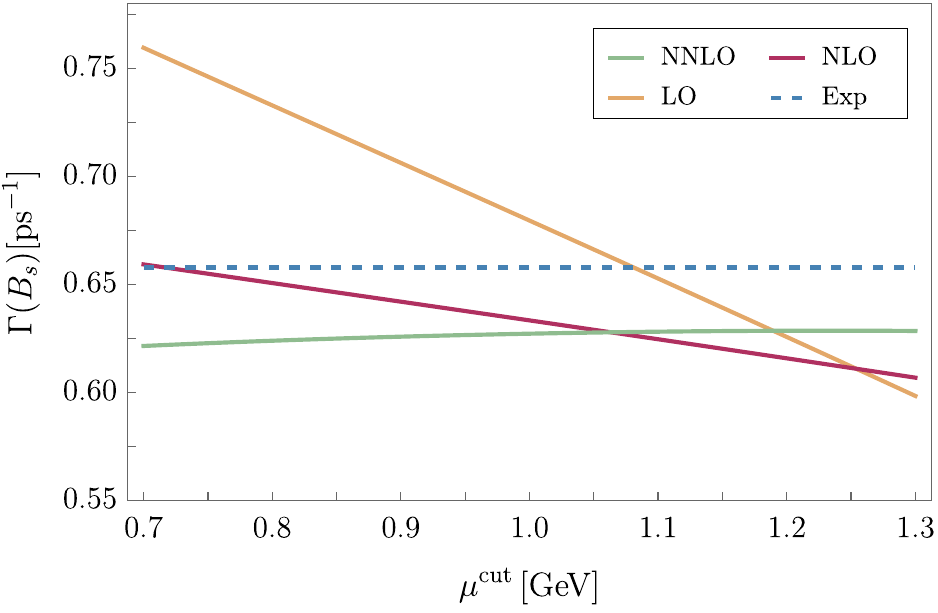} \quad
    \includegraphics[scale=0.42]{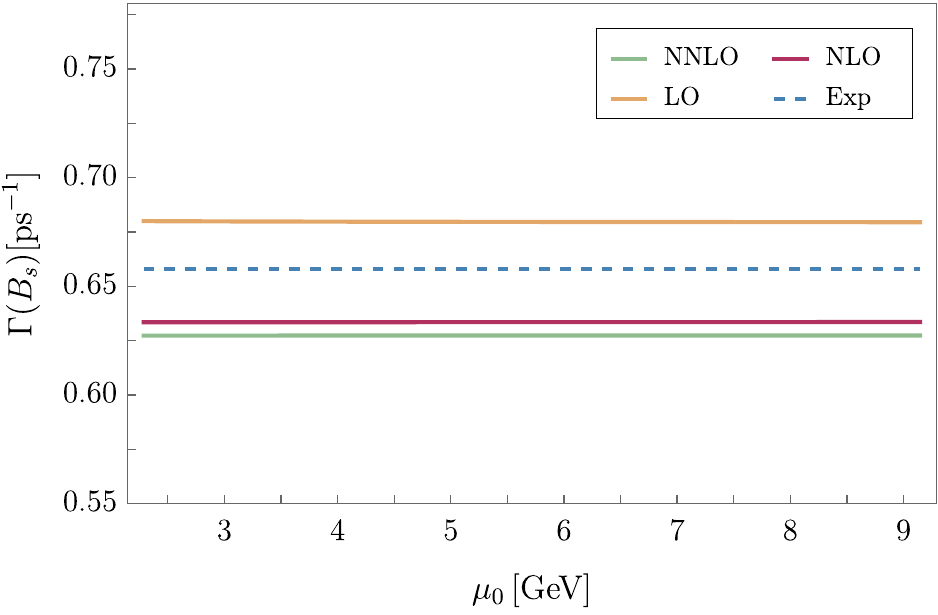}
    \caption{Dependence at NNLO- (solid green), NLO- (solid magenta), and LO-QCD (solid orange) of $\Gamma(B_s)$ on the renormalisation scales $\mu_b$ (top left), $\mu_c$ (top right), $\mu^{\rm cut}$ (bottom left), $\mu_0$ (bottom right), together with the corresponding experimental value (dashed blue).}
    \label{fig:Bs-kin-MS}
\end{figure}

In this Appendix we collect additional plots illustrating the dependence on the renormalisation scales $\mu_b, \mu_c, \mu^{\rm cut}$ and $\mu_0$ at LO-, NLO-, and NNLO-QCD, of the decay rates of the $B^+$, $B_d$, and $B_s$ mesons, shown respectively in Fig.~\ref{fig:Bu-kin-MS}, \ref{fig:Bd-kin-MS}, and \ref{fig:Bs-kin-MS}.


\newpage

\bibliographystyle{JHEP}
\bibliography{References}

\end{document}